%**start of header
\input amstex
\documentstyle{amsppt}
\magnification=\magstephalf
%%%%%%%%%%%% changes to amsppt.sty %%%%%%%%%%%%%%%%%%%%%%%
 \addto\tenpoint{\baselineskip 15pt
  \abovedisplayskip18pt plus4.5pt minus9pt
  \belowdisplayskip\abovedisplayskip
  \abovedisplayshortskip0pt plus4.5pt
  \belowdisplayshortskip10.5pt plus4.5pt minus6pt}\tenpoint
\pagewidth{6.5truein} \pageheight{8.9truein}
\subheadskip\bigskipamount
\belowheadskip\bigskipamount
\aboveheadskip=3\bigskipamount
\catcode`\@=11
\def\output@{\shipout\vbox{%
 \ifrunheads@ \makeheadline \pagebody
       \else \pagebody \fi \makefootline 
 }%
 \advancepageno \ifnum\outputpenalty>-\@MM\else\dosupereject\fi}
\outer\def\subhead#1\endsubhead{\par\penaltyandskip@{-100}\subheadskip
  \noindent{\subheadfont@\ignorespaces#1\unskip\endgraf}\removelastskip
  \nobreak\medskip\noindent}
\outer\def\enddocument{\par% \par will do a runaway check for \endref
  \add@missing\endRefs
  \add@missing\endroster \add@missing\endproclaim
  \add@missing\enddefinition
  \add@missing\enddemo \add@missing\endremark \add@missing\endexample
 \ifmonograph@ % do nothing
 \else
 \vfill
 \nobreak
 \thetranslator@
 \count@\z@ \loop\ifnum\count@<\addresscount@\advance\count@\@ne
 \csname address\number\count@\endcsname
 \csname email\number\count@\endcsname
 \repeat
\fi
 \supereject\end}
\def\endremark{\par\revert@envir\endremark\bigskip}
\catcode`\@=\active
%%%%%%%%%%%%%%% other macros %%%%%%%%%%%%%%%%%%%%%%%%%%%%
\CenteredTagsOnSplits
\NoBlackBoxes
\nologo
\def\today{\ifcase\month\or
 January\or February\or March\or April\or May\or June\or
 July\or August\or September\or October\or November\or December\fi
 \space\number\day, \number\year}
\define\({\left(}
\define\){\right)}
\define\Ahat{{\hat A}}
\define\Aut{\operatorname{Aut}}
\define\CC{{\Bbb C}}
\define\CP{{\Bbb C\Bbb P}}

\define\EE{\Bbb E}

\define\End{\operatorname{End}}

\define\HH{{\Bbb H}}
\define\Hom{\operatorname{Hom}}

\define\RR{{\Bbb R}}

\define\Spin{\operatorname{Spin}}

\define\ZZ{{\Bbb Z}}
\define\[{\left[}
\define\]{\right]}
\define\ch{\operatorname{ch}}
\define\chiup{\raise.5ex\hbox{$\chi$}}
\define\cir{S^1}

\define\dbar{{\bar\partial}}

%\define\exertag #1#2{\removelastskip\bigskip\medskip\eightpoint\noindent%
%\hbox{\rm\ignorespaces#2\unskip} #1.\ }  
\define\exertag #1#2{#2\ #1}

\define\inv{^{-1}}
\define\mstrut{^{\vphantom{1*\prime y}}}
\define\protag#1 #2{#2\ #1}
\define\rank{\operatorname{rank}}
\define\res#1{\negmedspace\bigm|_{#1}}
\define\temsquare{\raise3.5pt\hbox{\boxed{ }}}

\define\theprotag#1 #2{#2~#1}

\define\xca#1{\removelastskip\medskip\noindent{\smc%
#1\unskip.}\enspace\ignorespaces }

\define\zmod#1{\ZZ/#1\ZZ}

\define\zt{\zmod2}

\NoRunningHeads % USE IN FINAL VERSION; THEN COMMENT OUT NEXT LINE

\loadeusm
\define\Anom{\operatorname{Anomaly}}
\define\Bhat{\hat{B}}
\define\Bh{\check{B}}
\define\C#1{\Cal{C}(#1)}
\define\Cb#1{\overline{\C{#1}}}
\define\Ch{\check{C}}
\define\Cliff{\operatorname{Cliff}}
\define\Det{\operatorname{Det}}
\define\Dirac{D\hskip-.65em /}
\define\Ebar{\overline{\scrE}}
\define\Gbar{\overline{\Gamma }}
\define\Gtil{\tilde{G}}
\define\Hh{\check{H}}
\define\Htil{\tilde{H}}
\define\Kh{\check {K}}
\define\Lferm{L_{\text{fermi}}}
\define\MM{\Bbb M}
\define\Pfaff{\operatorname{Pfaff}}
\define\Qbar{\bQ}
\define\Qh{\check{Q}}
\define\RtpZ{\RR/2\pi \ZZ}
\define\SC{S_{\CC}}
\define\SSC{\eusm{S}_{\CC}}
\define\TCX{T_{\CC}X}
\define\TT{\Bbb{T}}
\define\VC{V_{\CC}}
\define\bQ{\overline{Q}}
\define\bul{\bullet}
\define\clf{\Omega _{\text{cl}}}
\define\ct{c^{(3)}}
\define\cyc#1{\scrF(#1)}
\define\dd{\bold{d}}
\define\gb{\bar{\gamma}}
\define\hchi{\hat\chi }
\define\hpsi{\hat\psi }
\define\image{\operatorname{image}}
\define\im{\operatorname{im}}
\define\jhat{\check {j}}
\define\pfaff{\operatorname{pfaff}}
\define\pnct{\EE^3\setminus \{0\}}
\define\pt{\operatorname{pt}}
\define\rtpz{\RR/\tpz}
\define\scrE{\Cal{E}}
\define\scrF{\Cal{F}}
\define\scrJ{\Cal{J}}
\define\scrP{\Cal{P}}
\define\spec{\operatorname{spec}}
\define\supp{\operatorname{supp}}
\define\tpz{2\pi \ZZ}
\define\triv#1{\bold{1}\mstrut _{#1}}
\define\voll{|d^nx|}

\input epsf

\define\conth#1{\bigskip#1\smallskip}
\define\contsh#1#2{\par	\indent \S#1.\enspace#2\endgraf}

\refstyle{A}
%\widestnumber\key{SSSSSSSSSSSSS}   % for widest bibliography name
\document
%**end of header

% \pretitle{$$\boxed{\boxed{\text{PRELIMINARY VERSION}}}$$\par\vskip 3pc}

	\topmatter
 \title\nofrills $K$-Theory in Quantum Field Theory \endtitle
 \author Daniel S. Freed  \endauthor
 \thanks The author is supported by NSF grant DMS-0072675.\endthanks
 \affil Department of Mathematics \\ University of Texas at Austin\endaffil 
 \address Department of Mathematics, University of Texas, Austin, TX
78712\endaddress 
 %\curraddr \endcurraddr
 \email dafr\@math.utexas.edu \endemail
 \date June 18, 2002 \enddate
 %\dedicatory \enddedicatory
 \abstract 
  We survey three different ways in which $K$-theory in all its forms enters
quantum field theory.  In Part 1 we give a general argument which relates
topological field theory in codimension two with twisted $K$-theory, and we
illustrate with some finite models.  Part 2 is a review of pfaffians of Dirac
operators, anomalies, and the relationship to differential $K$-theory.  Part
3 is a geometric exposition of Dirac charge quantization, which in
superstring theories also involves differential $K$-theory.  Parts 2 and 3
are related by the Green-Schwarz anomaly cancellation mechanism.  An
appendix, joint with Jerry Jenquin, treats the partition function of
Rarita-Schwinger fields.
 \endabstract
	\endtopmatter

\document

Grothendieck invented $K$-Theory almost 50~years ago in the context of
algebraic geometry, specifically in his generalization of the Hirzebruch
Riemann-Roch theorem~\cite{BS}.  Shortly thereafter, Atiyah and Hirzebruch
brought Grothendieck's ideas into topology~\cite{AH}, where they were applied
to a variety of problems.  Analysis entered after it was realized that the
symbol of an elliptic operator determines an element of $K$-theory.  Atiyah
and Singer then proved a formula for the index of such an operator (on a
compact manifold) in terms of the $K$-theory class of the symbol~\cite{AS1}.
Subsequently, $K$-theoretic ideas permeated other areas of linear analysis,
algebra, noncommutative geometry, etc.  One of the pleasant surprises of the
past few years has been the relevance of $K$-theory to superstring theory and
related parts of theoretical physics.  Furthermore, the story involves not
only topological $K$-theory, but also the $K$-theory of $C^*$-algebras, the
$K$-theory of sheaves, and other forms of $K$-theory.

Not surprisingly, this new arena for $K$-theory has inspired some
developments in mathematics which are the subject of ongoing research.  Our
exposition here aims to explain three different ways in which topological
$K$-theory appears in physics, and how this physics motivates the
mathematical ideas we are investigating.
 
Part~1 concerns {\it topological quantum field theory\/}.  Recall that an
$n$-dimensional topological theory assigns a complex number to every closed
oriented $n$-manifold and a complex vector space to every closed oriented
$(n-1)$-manifold.  Continuing the superposition principle and ideas of
locality to codimension two we are led to an extended notion which attaches
to each closed oriented $(n-2)$-manifold a special type of category, which we
term a {\it $K$-module\/}.  The~`$K$' stands for $K$-theory, and so by this
very general argument $K$-theory enters in codimension two, at least for
topological quantum field theories.  We illustrate these ideas with a finite
topological quantum field theory, a ``gauged $\sigma $-model'' which
generalizes the finite gauge theories studied in~\cite{F1}.  The $K$-modules
which enter here are a {\it twisted\/} form of ordinary $K$-theory.  It
should be noted that the appearance of twisted $K$-theory in the physics has
spurred its development in mathematics.\footnote{Twisted $K$-theory was
introduced in mathematics many years ago, both in topology~\cite{DK} and in
$C^*$-algebras~\cite{R}.  Recent references include~\cite{A2}
and~\cite{BCMMS}.  But a systematic development incorporating all of the
properties needed here has yet to be written.}  The 2-dimensional version of
the finite gauged $\sigma $-model was introduced in a different way
in~\cite{T2}.  Moore and Segal~\cite{M} make a general investigation of
``boundary states'' in 2-dimensional {\it topological\/} theories and are led
to this same finite gauged $\sigma $-model.  One observation here is that the
category of boundary states, at least in 2-dimensional topological theories,
is the $K$-module one encounters from general considerations in codimension
two.  From this point of view the appearance of $K$-theory is natural.
Turning to the 3-dimensional theory, once the pure gauge theory for finite
groups has been related to twisted $K$-theory---this connection was missed
in~\cite{F1}---it is natural to conjecture a similar relationship for compact
gauge groups of arbitrary dimension.  In particular, we identify the
``Verlinde algebra'' in Chern-Simons theory as a particular twisted
equivariant $K$-group.  This set of ideas is being pursued jointly with
M. Hopkins and C. Teleman.  (See~\cite{F3} for a more detailed motivational
account.)

Part~2 concerns {\it anomalies\/}; the main ideas go back to the mid 1980s.
We set the framework with a geometric picture of anomalies, and then explain
how the pfaffian line bundle of a family of Dirac operators encodes the
anomaly associated to the functional integral over a fermionic field.  The
detailed construction depends on the dimension~$n$ of the theory modulo~8.
One novelty here is the simultaneous presentation of all cases.
Unfortunately, its salient feature is the lack of a unified approach which
binds the different dimensions into a single picture---presumably held
together with Bott periodicity.  We have yet to find such a description.  The
topological anomaly is computed using the Atiyah-Singer index theorem, and it
is through the Atiyah-Singer formula that topological $K$-theory
enters.\footnote{There are notable exceptions in low dimensions, where the
topological anomaly is computed by a cohomological formula.}  In many cases,
however, the anomaly is geometric---it is a smooth line bundle with hermitian
metric and compatible connection.  With this motivation we are led to believe
that the geometric anomaly---the pfaffian line bundle with metric and
connection---may be computed using {\it differential $K$-theory\/}, a version
of $K$-theory which includes differential forms as ``curvatures''.  First
notions of differential $K$-theory appear in~\cite{Lo}, and a systematic
development of differential cohomology theories in general begins
in~\cite{HS}.  Ongoing joint work with M. Hopkins and I. Singer continues
these developments, in particular pursuing the connection between
differential $K$-theory and geometric invariants of families of Dirac
operators.  The particular connection with the pfaffian is what is needed for
anomalies.
 
Part~3 is an elementary exposition of {\it Dirac charge quantization\/}.  In
classical electromagnetism, and its generalizations in supergravity with
forms of higher degree, charges take values in real cohomology.  In quantum
theories charges are constrained to lie in a full lattice inside real
cohomology, and from a mathematical point of view it is natural that the
appropriate lattice is determined by a generalized cohomology theory.
Ordinary integral cohomology is the traditional choice, but in superstring
theory it is $K$-theory---in some cases the real or quaternionic
version---which has proved relevant.  The electric and magnetic currents in
these theories must simultaneously encode local information---the positions
and velocities of charges---as well as the global information of charge
quantization.  The geometric objects which accomplish this are cocycles for
generalized differential cohomology theories, e.g., for differential
$K$-theory.  This provides further impetus for the development of
differential cocycles.  We emphasize the easy examples, where the geometry of
charge quantization is more readily accessible.  One important point is the
anomaly in the electric coupling if there is simultaneous magnetic and
electric current.  We conclude Part~3 with some remarks explaining why
$K$-theory quantizes Ramond-Ramond charge in superstring theory.
 
The occurrences of $K$-theory in Parts~2 and~3 are related.  Namely, the
anomaly in the electric coupling can cancel anomalies from fermions if the
former may be computed in differential $K$-theory since, as explained above,
the latter are conjecturally computed in differential $K$-theory using a
refinement of the Atiyah-Singer index theorem.  Therefore, the main
ingredient in the {\it Green-Schwarz anomaly cancellation
mechanism\/}~\cite{GS}, extended to include global as well as local
anomalies, is a geometric form of the Atiyah-Singer index theorem for
families of Dirac operators.  The superstring examples are explained from
this point of view in~\cite{FH}, ~\cite{F5}.  In ongoing work with J. Distler
we consider generalizations in superstring theory, and with E. Diaconescu and
G. Moore we are investigating similar questions in M-theory.
 
There is an appendix, joint with Jerry Jenquin, which is a pedagogical
account of Rarita-Schwinger fields and their quantization to obtain
spin~$3/2$ particles.  The anomaly computation for these fields is a bit
confusing, as it typically looks different in odd and even dimensions.  Our
treatment is uniform for all dimensions, chiralities, and multiplicities.

A detailed table of contents:

{\narrower\narrower\narrower
 \conth{Part 1: Topological Quantum Field Theory in Codimension Two}
 \contsh{1.1}{General Remarks}
 \contsh{1.2}{The Groupoid of Fields in Finite TQFT}
 \contsh{1.3}{Extended TQFTs from Functional Integrals}
 \contsh{1.4}{Finite TQFT}
 \contsh{1.5}{The One-Dimensional Theory}
 \contsh{1.6}{Twisted $K$-Theory}
 \contsh{1.7}{The Two-Dimensional Theory}
 \contsh{1.8}{The Three-Dimensional Theory}
 \conth{Part 2: Anomalies and Pfaffians of Dirac Operators}
 \contsh{2.1}{Actions in Euclidean QFT}
 \contsh{2.2}{The Partition Function of a Spinor Field}
 \contsh{2.3}{Construction of~$\Pfaff D$}
 \contsh{2.4}{Computation of the Topological Anomaly}
 \contsh{2.5}{Computation of the Geometric Anomaly}
 \conth{Part 3: Abelian Gauge Fields}
 \contsh{3.1}{Maxwell's Equations}
 \contsh{3.2}{The Action Principle for Electromagnetism}
 \contsh{3.3}{Dirac Charge Quantization}
 \contsh{3.4}{The Electric Coupling Anomaly}
 \contsh{3.5}{Generalized Differential Cocycles}
 \contsh{3.6}{Self-Dual Fields}
 \contsh{3.7}{Ramond-Ramond Charge and $K$-Theory}
 \conth{Appendix: The Partition Function of Rarita-Schwinger Fields}
 \contsh{A.1}{Quantization of Spinor Fields: Review}
 \contsh{A.2}{Quantization of Rarita-Schwinger Fields: First Approach}
 \contsh{A.3}{Quantization of Rarita-Schwinger Fields: Second Approach}
 \contsh{A.4}{Quantization of Rarita-Schwinger Fields: Third Approach}
 \contsh{A.5}{The Euclidean Partition Function of a Rarita-Schwinger Field}
 \contsh{A.6}{Two Illustrative Examples}
}\bigskip

I have discussed these topics with many people over a long period.  I
particularly thank my recent collaborators Emanuel Diaconescu, Jacques
Distler, Jerry Jenquin, Mike Hopkins, Greg Moore, Is Singer, Constantin
Teleman, and Ed Witten.

 \newpage
 \head
 Part 1: Topological Quantum Field Theory in Codimension Two
 \endhead
 \comment
 lasteqno 1@ 14
 \endcomment

 \subhead \S{1.1}. General Remarks
 \endsubhead

Some essential features of a topological quantum field theory (TQFT) were
abstracted in~\cite{A1}; refinements were discussed in~\cite{Q}, \cite{T1},
\cite{Wa} and other references as well.  A TQFT gives algebro-topological
invariants of manifolds, but unlike typical homotopical invariants is tied to
a specific dimension~$n$ and may depend on the differentiable structure.  The
basic data is the {\it partition function\/}
  $$ X^n \longmapsto Z(X^n) \tag{1.1} $$
which assigns to every compact oriented $n$-manifold a complex number~$Z(X)$.
This is a topological invariant in the sense that the invariants of {\it
diffeomorphic\/} manifolds agree.  Also, if $-X$ denotes the oppositely
oriented manifold, then $Z(-X)= \overline{Z(X)}$.  The quantum nature of~$Z$
manifests itself in its multiplicative properties.  For example, the
invariant of a disjoint union is the product of the invariants of the
constituent manifolds:
  $$ Z(X_1 \sqcup X_2) = Z(X_1)\,Z(X_2). \tag{1.2} $$
This generalizes to a {\it gluing law\/} when a closed manifold~$X$ is cut
along an oriented hypersurface~$Y$.  Assume for simplicity that $Y$
separates~$X$ into two pieces~$X_1$ and~$X_2$, as in Figure~1.  Then the {\it
local\/} nature of quantum field theory predicts that there should be an
invariant~$Z(X_i)$ defined for the manifolds~$X_i$, which now have a nonempty
boundary, and that a multiplicative equation analogous to~\thetag{1.2} should
hold.  In fact, field theory on a manifold with boundary requires boundary
conditions for the fields, which leads not to a single complex-valued
invariant, but rather a more complicated invariant which is in some sense a
function of boundary conditions.  The linearity of quantum mechanics leads us
to suspect that the invariant is a vector 
  $$ Z(X_i) = \bigl(Z(X_i)_1,\dots ,Z(X_i)_N \bigr), \tag{1.3} $$
where $Z(X_i)_j\in \CC$ and the indices depend only on the boundary~$Y$.  In
typical quantum field theories $N=\infty $, of course, but in special
topological theories, such as the ones considered here, it is finite.  We may
go further and postulate an assignment
  $$ Y^{n-1} \longmapsto E(Y) \tag{1.4} $$
of a Hilbert space~$E(Y)$ to every closed oriented $(n-1)$-manifold---whether
or not it is a boundary---and an assignment
  $$ X^n \longmapsto Z(X)\in E(\partial X) \tag{1.5} $$
to every compact oriented $n$-manifold of a vector in the Hilbert space of
the boundary.  Furthermore, \thetag{1.4}~should be functorial in the sense
that a diffeomorphism $Y' \to Y$ induces an isomorphism $E(Y') \to E(Y)$.
Finally, the gluing law which generalizes~\thetag{1.2} in the situation of
Figure~1 is 
  $$ Z(X) = \langle Z(X_1),Z(X_2)  \rangle_{E(Y)}. \tag{1.6} $$

 \midinsert
 \bigskip
 \centerline{
 \epsffile{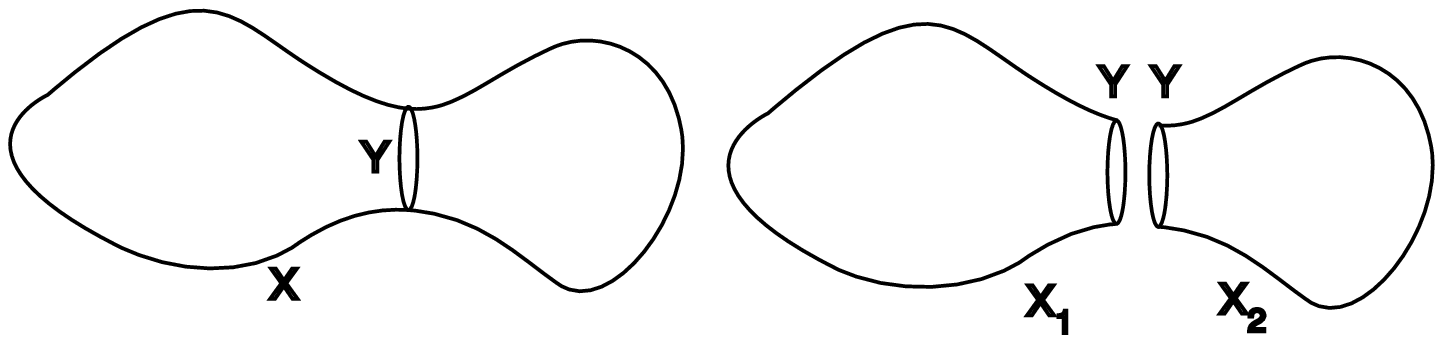}}
 \nobreak
 \botcaption{Figure~1: Cutting a closed manifold in codimension one}
 \endcaption
 \bigskip
 \endinsert

These are the highlights of the standard axioms, but it is tempting to go
further.  Namely, of the two basic properties of the partition
function~\thetag{1.1}---functoriality and locality---we have only imposed
functoriality on the invariant~$Y\mapsto E(Y)$ in~\thetag{1.5}; it is
tempting to impose locality as well.  Thus if~$Z^{n-2}\subset Y^{n-1}$ splits
the closed oriented manifold~$Y$ into a union~$Y_1 \cup Y_2$, we would like
to factorize the vector space~$E(Y)$ as some sort of product of invariants
associated to the~$Y_i$, which are manifolds with boundary.  Now
in~\thetag{1.3} we can consider $Z(X_i)$~as a vector of complex numbers, so
by analogy we expect that $E(Y_i)$~should be a vector of Hilbert spaces.
Further, we then expect an assignment
  $$ Z^{n-2} \longmapsto \scrE(Z)  $$
which attaches to a closed oriented $(n-2)$-manifold a ``module'' over the
``ring'' of Hilbert spaces.  This module should have an ``inner product''
back to the ground ``ring'', and then the gluing law analogous
to~\thetag{1.6} should state
  $$ Z(Y) = \langle Z(Y_1),Z(Y_2)  \rangle_{E(Z)}.   $$
This extended notion of a TQFT can be made precise in various ways and occurs
in different forms in the literature (e.g.~\cite{L}).  We content ourselves
here with a heuristic description as follows.

The collection of all finite dimensional Hilbert spaces forms a tensor
category using the operations of direct sum and tensor product.  The
classifying space of this category is a product of classifying spaces for the
general linear groups of all dimensions.  Apparently what is needed here
instead is a tensor category ~$K$ whose classifying space is a classifying
space for complex $K$-theory, i.e., is homotopic to~$\ZZ\times BU$.  So
$E(Y)$~is a category with an action of~$K$, i.e., a {\it $K$-module\/}.  We
will use the category of finite dimensional Hilbert spaces as a heuristic
for~$K$, but a more careful account would substitute a correct model for~$K$
instead.  Also, we will not always mention the inner product in $K$-modules,
even though it exists in unitary theories and the finite theories we
construct are unitary.
 
Summary: A TQFT assigns to a closed oriented manifold a complex number in the
top dimension, a complex vector space in codimension one, and a $K$-module in
codimension~2.

 \subhead \S{1.2}. The Groupoid of Fields in Finite TQFT
 \endsubhead

Let $G$~be a finite group and $S$~a finite $G$-set, that is, a finite set
with a $G$-action.\footnote{It suffices to restrict to transitive $G$-sets,
that is, $S=G/H$ for a subgroup~$H\subset G$, since any $G$-set decomposes
into a disjoint union of such and the construction of TQFTs decomposes
accordingly.  But the analogy to continuous theories is clearer in the
framework of arbitrary~$S$.}  Let $M$~be a topological space, which we will
soon specialize to be a compact manifold.  Let $\C M$~be the category whose
objects are pairs~$(P,\phi )$, where $P\to M$ is a principal $G$-bundle
(Galois covering) and $\phi \:P\to S$ is a $G$-equivariant map.  A morphism
$f\:(P',\phi ')\to (P,\phi )$ is a $G$-equivariant map~$f\:P'\to P$ which
commutes with the projections to~$S$ and such that $\phi ' = f\circ \phi $.
It is easy to see that every morphism is invertible: $\C M$~is a groupoid.
Let $\Cb M$~be the set of equivalence classes of objects in~$\C M$.  It is
finite if $M$~is compact.
 
The groupoid~$\C M$ is the collection of {\it fields\/} on~$M$.  A model with
this set of fields is called a {\it gauged $\sigma $-model\/}.  It is usually
defined for $G$~a Lie group and $P,M,S$ smooth manifolds.  If $G$~has
positive dimension, then an object in~$\C M$ also includes a connection
on~$P\to M$ and morphisms are required to pullback connections.  For $G=\{1\}$
the collection of fields~ $\C M$ is the set of maps $\phi \:M\to S$, and the
model is a $\sigma $-model with target~$S$.  If $S$~is a point with trivial
$G$-action, then we have a pure gauge theory with gauge group~$G$.  In the
general case the $G$~action on~$S$ determines a bundle $S_P \to M$ with
typical fiber~$S$ associated to a principal $G$-bundle $P\to M$, and the
map~$\phi $ may be viewed as a section of this associated bundle.

It is convenient to replace~$\C M$ by an equivalent category.  Fix a smooth
universal $G$-bundle $EG\to BG$.\footnote{For example, embed $G$ in a unitary
group~$U(N)$ and take~$EG$ to be the Stiefel manifold of unitary maps of
$\CC^N$ into a complex Hilbert space.  In theories with~$\dim G>0$ we also
fix a connection on the universal bundle.}  There is a category of
triples~$(P,\gamma ,\phi )$, where $P,\phi $~are as before and $\gamma \:P\to
EG$ is a $G$-equivariant map; the quotient map $\gb\:M\to BG$ is a
classifying map for~$P$.  A morphism $f:(P',\gamma ', \phi ')\to
(P,\gamma ,\phi )$ is a map $f\:P'\to P$ where, as above, $\phi '=f\circ \phi
$.  Note that there is no condition on the classifying maps.  The forgetful
functor $(P,\gamma ,\phi ) \rightsquigarrow (P,\phi )$ is an equivalence of
categories. 
 
For computations it is convenient to use a much smaller equivalent groupoid.
For example, when $M$~is a point we have
  $$ \C {\pt} \approx \text{$S$ as a $G$-set}. \tag{1.7} $$
Recall that attached to any $G$-set~$S$ is a groupoid: the set of objects
is~$S$ and each pair~$(g,s)\in G\times S$ corresponds to a morphism $s\to
g\cdot s$.  Consider next $M=S^1$.  We rigidify~$\C \cir$ by fixing a
basepoint~$*\in \cir$ and requiring that $G$-bundles $P\to \cir$ have a
basepoint in the fiber over~$*$.  Such a based bundle is determined up to
unique isomorphism by its holonomy, which is an element of~$G$.  The
map~$\phi \:P\to S$ is determined by its value at the basepoint of~$P$, which
must be fixed by the action of the holonomy.  So
  $$ \C\cir \approx \text{$G$~acting on $\{(s,g): s\in S, g\in G, g\cdot
     s=s\}$} \tag{1.8} $$
The action of~$G$ translates the basepoint in the fiber over~$*$.  The
morphism which corresponds to~$h\in G$ maps $(s,g)\mapsto (h\cdot s, hgh\inv
)$.

 \subhead \S{1.3}. Extended TQFTs from Functional Integrals
 \endsubhead

This general discussion is taken from~\cite{F1}, where it was applied to
finite theories with~$S=\pt$.

In an $n$-dimensional quantum field theory the exponentiated classical action
on a compact oriented $n$-manifold~$X$ is a function\footnote{When we discuss
anomalies in Part~2, we will see that in general quantum field theories the
action is an element of a hermitian line, but in finite TQFT there is no
anomaly.}
  $$ e^{iS_X}\:\Cb X\:\longrightarrow \CC. \tag{1.9} $$
For finite theories, where $\Cb X$~is a finite set, a measure is simply a
function $\Cb X\to\RR^{>0}$.  Given a measure and classical action, the
quantum partition function is the integral 
  $$ Z(X) = \int_{\Cb X}e^{iS_X(\varphi )} \;d\varphi . \tag{1.10} $$
For a finite theory \thetag{1.10}~reduces to a finite sum.
  
The classical action~\thetag{1.9} extends to compact oriented $n$-manifolds
with boundary, but it is sometimes not valued in the complex numbers.
Rather, in general its value on a field~$\varphi $ lies in a hermitian line
which depends only on the restriction of~$\varphi $ to~$\partial X$.  It is
natural, then to extend the notion of classical action to codimension one.
That is, given a compact oriented $(n-1)$-manifold~$Y$ we assign to every
field on~$Y$ a hermitian line.  Recall that the fields~$\C Y$ form a
groupoid, and so we require that morphisms act as isomorphisms on the
hermitian lines.  In other words, the action gives an equivariant hermitian
line bundle over~$\C Y$.  The quantum Hilbert space~$E(Y)$ is usually
described as the vector space of invariant sections of this line bundle, with
an $L^2$~metric relative to a measure on $\Cb Y$.  (In~\cite{F1} it is also
described as the result of integrating the classical action over~$\Cb Y$.)
There is then a natural extension of~\thetag{1.10} to manifolds with
boundary.
 
The story proceeds analogously in codimension two.  In codimension one we can
say that the classical action takes values in one-dimensional $\CC$-modules.
The extension to a compact oriented $(n-2)$-manifold~$Z$ assigns to every
field a one-dimensional $K$-module, which we term a {\it
$K$-line\/}.\footnote{In a unitary theory the $K$-lines have a ``hermitian
structure'' but we will not be careful about it.  In~\cite{F2} $K$-modules
(of arbitrary dimension) with hermitian structure are called ``2-Hilbert
spaces''.}  Again morphisms in the groupoid~$\C Z$ act in a natural way, and
$\scrE(Z)$~is defined to be the $K$-module of ``invariant'' sections.  As the
quotes indicate, we must be careful to interpret the notion of invariance
correctly.  In codimension one, where the classical action on~$Y$ is a line
bundle over~$\C Y$, the set of invariants in the fiber at a field~$\varphi $
is either the entire fiber or zero, according as the action of the
automorphism group at~$\varphi $ is trivial or not.  But in codimension two
the fiber is a category, not a set, so it is natural to interpret invariance
under the automorphism group as being a representation of the automorphism
group on the fiber.  Thus if we trivialize the $K$-line at a particular
field, the invariants under the automorphism group form the category of
(finite dimensional) representations of the automorphism group.
 
This construction of a TQFT from an extended notion of a classical action
allows us to deduce formally the axioms of the quantum theory from simple
properties of the classical action and the measure.

 \subhead \S{1.4}. Finite TQFT
 \endsubhead

Recall that the groupoids of fields in finite TQFT are determined by a finite
group~$G$ and a finite $G$-set~$S$.  On any compact space~$M$ the groupoid of
fields~$\C M$ is discrete with finite automorphism groups, and on such a
groupoid there is a natural {\it counting measure\/}:  To each
object~$\varphi \in \C M$ we assign the positive number 
  $$ \mu _M(\varphi ) = \frac{1}{\#\Aut\varphi }. \tag{1.11} $$
The counting measure clearly descends to~$\Cb M$. 
 
The classical action in the $n$-dimensional finite TQFT is determined by a
cocycle~$B$ for an element in the equivariant cohomology
group~$H^n_G(S;\RR/2\pi \ZZ)$.  Concretely, we fix a singular cocycle of
degree~$n$ with coefficients in~$\RtpZ$ on the total space of the bundle
$S_{EG}\to BG$.  The special case~ $B=0$ is already interesting and does not
require the formalism explained in the next paragraph.  Only the cohomology
class of~$B$ matters in the sense that an $(n-1)$-cochain~$A$ determines an
isomorphism between the theory defined by~$B$ and the theory defined by~$B +
\delta A$.  Notice, since $G$~and $S$~are finite, that $H^n_G(S;\RtpZ)\cong
H^{n+1}_G(S;\ZZ)$ for $n\ge1$.

To write the classical action we integrate singular cocycles over compact
oriented manifolds.  What we need appears in~\cite{F1,Appendix}, and as we
explain later it is a special case of a more general integration theory for
differential cocycles~\cite{HS}.\footnote{The theory of differential cocycles
is one context in which ``$K$-line'' and its generalizations acquire a
precise mathematical meaning.}  We summarize briefly what we need.  Let
$M$~be a compact oriented manifold and $b$~a singular cocycle of degree~$n$
with coefficients in ~$\RtpZ$.  In the simplest case, if $M=X$~is closed (no
boundary) of dimension~$n$, then there is a cohomological pairing of the
cohomology class of~$b$ with the fundamental class~$[X]$ of~$X$ which
produces an element of~$\RR/2\pi \ZZ$.  This is the integral of~$b$ over~$X$.
It is defined by evaluating~$b$ on any cycle representing~$[X]$.  Now suppose
instead that $M=Y$~is closed of dimension~$(n-1)$.  Then of course there is
no cohomological meaning to the integral of~$b$ over~$Y$.  Instead, we claim
that this integral may be interpreted as a hermitian line as follows.  Let
$\cyc Y$ be the category whose objects are singular $(n-1)$-cycles which
represent the fundamental class~$[Y]$.  Let a singular $n$-chain~$y$
determine a morphism $x\to x+\delta y$ for all~$x$.  Hence $\cyc Y$~is a
groupoid which is connected---there is a morphism between any two objects.
Over~$\cyc Y$ we consider a hermitian line bundle whose fiber at each object
is~$\CC$ with its canonical metric, and such that a morphism defined by~$y$
acts as multiplication by~$\exp(i\langle b,y \rangle)$, where $\langle \cdot
,\cdot \rangle$~is the pairing of cochains and chains.  The hermitian line~
$\exp(i\int_{Y}b)$ is the hermitian line of invariant sections of this
hermitian line bundle over~$\cyc Y$.  An analogous construction defines a
$K$-line ~$\exp(i\int_{Z}b)$ for $M=Z$~a closed $(n-2)$-manifold.  Namely, we
attach the trivial $K$-line to each object of~$\cyc Z$ and to each morphism
use a slight generalization of the previous construction to attach a
hermitian line to each morphism in~$\cyc Z$.  These lines act on~$K$ by
multiplication (tensor product).  Then $\exp(i\int_{Z}b)$~is the $K$-line of
invariant sections.  There are similar constructions for manifolds with
boundary using cycles which represent the relative fundamental class.
 
Fix a cocycle~$B$ on~$S_{EG}$ as above.  To construct the classical action of
finite TQFT on a compact oriented manifold~$M$ of dimension~$\le n$, we use
the category~$\C M$ of triples~$(P,\gamma ,\phi )$, where $P\to M$ is a
principal $G$-bundle, $\gamma \:P\to EG$ is a $G$-equivariant map, and $\phi
\:M\to S_P$ is a section.  Let $\gamma _S\:S_P\to S_{EG}$ be the classifying
map associated to~$\gamma $.  Then set
  $$ e^{iS_M(\varphi )} = \exp\Bigl(i\int_{M}\phi ^*\gamma
     _S^*B\Bigr).  $$
As explained in the previous paragraph for $M$~closed of dimension~$n$ this
is a complex number, for $M$~closed of dimension~$n-1$ it is a hermitian
line, etc. 
 
Appropriate functoriality and gluing laws for the classical action follow
from elementary facts about chains and cochains.  It follows that
\thetag{1.10}~and the constructions which follow yield an extended TQFT.
This follows the physicists' paradigm of the functional integral with two
notable exceptions: (i)\ we have extended the construction down to
codimension two, and (ii)\ the functional integrals here are over a finite
measure space, so reduce to finite sums.  We now analyze the theory for
$n=1,2,3$.

 \subhead \S{1.5}. The One-Dimensional Theory
 \endsubhead

In general a one-dimensional TQFT attaches a Hilbert space~$E=E(\pt)$ to a
point, the identity map to the closed interval, and the complex number~$\dim
E$ to a circle.  So we need only identify~$E$.  Recall that $B$~is a cocycle
for~$H^1_G(S;\RtpZ)\cong H^2_G(S;\ZZ)$.  There is a natural equivalence of
categories between such cocycles and the category of $G$-equivariant
hermitian line bundles over~$S$, so we fix such a hermitian line bundle $L\to
S$.  Recall from~\thetag{1.7} that the groupoid of fields on~$\pt$ is
equivalent to~$S$ with its $G$-action, so it follows that $E$~is the space of
invariant sections of $L\to S$.  The $L^2$~metric is determined by the
hermitian metric on~$L$ and the measure on~$S$ which assigns to~$s\in S$ the
inverse of the cardinality of the stabilizer subgroup of~$s$.
 
In general the finite TQFT splits as a sum of theories over the orbits of the
$G$-action on~$S$.  For a transitive $G$-space~$S=G/H$ the cocycle which
defines the theory is a character of~$H$, and the Hilbert space is zero
unless the character is trivial, in which case it is one-dimensional.  It is
instructive to compute the partition function of~$\cir$ in this case
directly.  One can identify $\Cb\cir\approx H$, and for the theory defined by
the character~$\chi $ \thetag{1.10}~reduces to
  $$ Z(\cir) \;=\; \sum\limits_{h\in H}\chi (h)\;=\;\cases \#H ,&\chi
     ~\text{trivial};\\0,&\text{otherwise}.\endcases  $$

There is a variation of this theory where in addition to the cocycle~$B$ we
also choose a cocycle for an element of~$H^0_G(S;\zt)$.  This amounts to
replacing the line bundle over~$S$ with a {\it $\zt$-graded line bundle\/}.
The resulting quantum theory has a $\zt$-graded Hilbert space.  There are
corresponding generalizations in~$n=2,3$ which we will not pursue here.

 \subhead \S{1.6}. Twisted $K$-Theory
 \endsubhead

The theories in $n=2,3$~dimensions involve $K$-theory in codimension two, as
explained earlier, and the cocycle~$B$ leads to a twisted form of
$K$-theory.  We digress to introduce twisted $K$-theory.

Twistings are common in ordinary cohomology.  For example, a flat real line
bundle over a manifold~$T$ determines a twisting of its real cohomology.  It
can be easily described in de Rham theory.  Similarly, a local system over a
space twists its integral cohomology.  In algebro-topological terms one can
define twistings of any generalized cohomology theory.  Briefly, such a
theory is defined by a spectrum~$R$; the cohomology groups of a space~$T$ are
the sets of homotopy classes of maps from~$T$ into loop spaces of~$R$.  A
one-dimensional twisting over~$T$ is a cocycle for
$H^1\bigl(T;GL_1(R)\bigr)$, and this leads to a fibration of spectra over~$T$
with typical fiber~$R$.  Now if $R$~is a {\it ringed\/} spectrum, i.e., it
leads to a {\it multiplicative\/} cohomology theory, then $GL_1(R)$~is the
group of units in~$R$, which is interpreted as a space, more precisely an
$H$-group.  One can deduce the units~$GL_1(K)$ of $K$-theory from results
in~\cite{DK}, \cite{S1}, \cite{ASe}, \cite{AP}.  The relevant units for our
purposes comprise the Eilenberg-MacLane space~$K(\ZZ,2)\sim\CP^{\infty}$.
Intuitively, if we think of $K$-theory as the category of finite-dimensional
complex vector spaces,\footnote{As we pointed out earlier, the classifying
space of this category is {\it not\/} that of $K$-theory.  Nonetheless, it is
useful as a heuristic approximation.} then the subcategory of complex
lines---that is, one-dimensional complex vector spaces---is the collection of
multiplicative units, where multiplication is by tensor product.\footnote{If
we pursue the variation mentioned at the end of the previous section, then we
encounter another type of twisting classified on a space~$T$ by~$H^1(T;\zt)$.
The corresponding group of all units---including those which lead
to~$K(\ZZ,2)$ above---is heuristically the subcategory of $\zt$-graded lines
in the category of $\zt$-graded complex vector spaces, where we use the
latter as a heuristic model for~$K$.  There are other units in $K$-theory
which have not (yet?) appeared in quantum field theory.}  These twistings of
$K$-theory over a space~$T$ are classified up to isomorphism by~$H^3(T;\ZZ)$.

The twisted $K$-theory depends on a choice of {\it cocycle\/}~$B$, not just a
cohomology class.  We use the notation $K^{\bul+B}(T)$ to denote the
$K$-theory groups of~$T$ twisted by~$B$.  Isomorphisms of cocycles lead to
isomorphisms of the corresponding twisted $K$-theory groups.  Thus,
$H^2(T;\ZZ)$~acts as automorphisms on a twisted $K$-theory
group~$K^{\bul+B}(T)$.  More generally, $K^{\bul+B}(T)$ is a ($\ZZ$-graded)
module over~$K^\bul(T)$; the action of $H^2(T;\ZZ)$ is via its image
in~$K^\bul(T)$ realizing a degree two cohomology class as a complex line
bundle.
 
The space of all sections of the bundle of spectra over~$T$ determined by~$B$
is a $K$-module whose Grothendieck group is twisted $K$-theory.  For the
application to finite TQFT the space~$T$ is a finite set and we will describe
such $K$-modules in different terms.  The $K$-theory spectrum has concrete
realizations, for example as spaces of Fredholm operators, and this leads to
more concrete pictures of twisted $K$-theory~\cite{A2}, \cite{BCMMS}.
 
We encounter twistings of {\it equivariant\/} $K$-theory as well.  Recall
that for a compact Lie group~$G$ the equivariant group~$K_G^0(\pt)$ is
isomorphic to the representation ring of~$G$; the odd equivariant $K$-theory
of a point vanishes.  The category of cocycles for the equivariant cohomology
group~$H^3_G(\pt)$ is equivalent to the category of central
extensions~$\Gtil$ of~$G$ by the circle group~$\TT$:
  $$ 1 \longrightarrow \TT\longrightarrow \Gtil\longrightarrow
     G\longrightarrow 1.  $$
The twisted $K$-group~$K^{0+\Gtil}_G(\pt)$ is then the set of equivalence
classes of finite-dimensional $\zt$-graded complex representations of~$\Gtil$
on which the center~$\TT$ acts by the standard representation.  It is a
module over~$R(G)$.  A heuristic model for~$K_G$ is the category of all
finite-dimensional $\zt$-graded representations of~$G$.  Then the category of
finite-dimensional $\zt$-graded representations of~$\Gtil$ on which the
center is standard is a $K_G$-module whose Grothendieck group is the twisted
equivariant $K$-theory.

 \subhead \S{1.7}. The Two-Dimensional Theory
 \endsubhead

The Hilbert space~$E=E(\cir)$ attached to the circle in a two-dimensional
TQFT is a commutative associative algebra with unit and inner product such
that the trilinear form $x,y,z\mapsto \langle xy,z \rangle$ is totally
symmetric.  This structure, called a {\it Frobenius algebra\/}, is derived
from the functional integral over simple surfaces---spheres with disks
removed.  There are many accounts in the literature; \cite{D}~ is one of the
earliest.  For mathematical accounts, see ~\cite{Sa}, \cite{Ab}.  If this
algebra is semisimple, as it is in the examples we encounter, it can be
diagonalized and then there is an explicit formula for the partition function
of an oriented surface.  It is also true that the field theory is determined
by the Frobenius algebra.  An extended two-dimensional TQFT, of the type we
are considering, has a more refined invariant: the
$K$-module~$\scrE=\scrE(\pt)$ attached to a point.  Then $E$~may be computed
in terms of~$\scrE$ as
  $$ E \cong \overline{\scrE}\otimes \CC, \tag{1.12} $$
where $\overline{\scrE}$~is the Grothendieck group of~$\scrE$.  Although
\thetag{1.12}~is a $\CC$-algebra, there is no natural $K$-algebra structure
on~$\scrE$.  In particular, there is not necessarily a ring structure
on~$\overline{\scrE}$ which induces the $\CC$-algebra structure on~$E$. 
 
The starting data for finite TQFT in two dimensions is a cocycle~$B$ for
$H^2_G(S;\RtpZ)\cong H^3_G(S;\ZZ)$.  As explained above this gives a twisting
of the $G$-equivariant $K$-theory of~$S$.  We can interpret $B$~as providing
a $K$-line attached to each point of~$S$ together with a lifting of the
$G$-action, i.e., an $G$-equivariant $K$-line bundle over~$S$.  Indeed, this
must be so to be consistent with~\thetag{1.7}.  Note that for~$S=\pt$ the
cocycle~$B$ assigns a hermitian line to each~$g\in G$, since lines are the
units in~$K$, and the action determines a group law on the union of the
lines, which is a central extension~$\Gtil$ of~$G$ by~$\TT$.  The
relationship between integral~ $H^3$ and $\TT$-gerbes is explained in many
recent works, e.g.~\cite{B} and ~\cite{H}.  On the other hand $\TT$-gerbes
are equivalent to $K$-lines, just as $\TT$-torsors are equivalent to
(hermitian) $\CC$-lines.
 
The $K$-module~$\scrE$ is the space of invariant sections of this $K$-line
bundle.  Recall the notion of ``invariant section'' explained earlier.  At a
point of~$S$ the action of the stabilizer group~$H\subset G$ determines a
central extension~$\Htil$ of~$H$ by~$\TT$, and the $K$-module of invariant
sections at that point is the category of representations of~$\Htil$ on which
the center acts by scalar multiplication.  We can describe an object
in~$\scrE$ more explicitly if, following~\cite{F1,\S8}, we identify the
$K$-line attached to each point of~$S$ with the trivial $K$-line.  Then the
action of~$G$ on the trivialized $K$-line bundle over~$S$ attaches a
$\CC$-line~$L_{(s,g)}$ to every~$s\in S,\,g\in G$.  It acts by tensor product
as a morphism from the trivial $K$-line at~$s$ to the trivial $K$-line
at~$g\cdot s$, and there is a composition law for the action.  An object
of~$\scrE$ is then a $\zt$-graded complex vector space~$W_s$ attached to
each~$s\in S$ together with an isomorphism $L_{(s,g)}\otimes W_s\to W_{g\cdot
s}$ for every~$s\in S,\,g\in G$.  For the theory with $B=0$ this specializes
to the category of $G$-equivariant $\zt$-graded complex vector bundles
over~$S$.

The Grothendieck group~$\Ebar$ of~$\scrE$ is the twisted equivariant
$K$-theory group~$K_G^{0+B}(S)$.

The Hilbert space~$E$ attached to~$\cir$ may be computed directly from the
description~\thetag{1.8} of the groupoid of fields on~$\cir$.  Note that the
action attaches a hermitian line~$L_{(s,g)}$ to each pair~$(s,g)$
with~$g\cdot s=s$; this line is a special case of the hermitian lines
considered above, and since $s$~is fixed by~$g$ it does not depend on a
trivialization of the $K$-line at~$s$.  The group~$G$ acts on the
set~$\scrP$ of pairs~$(s,g)$ with $g\cdot s=s$, and $E$~is the space of
invariant sections of the line bundle~$L\to\scrP$.  Note that $\scrP$~is a
groupoid, and therefore there is a natural product on
  $$ E\cong \Bigl[\bigoplus\limits_{(s,g):g\cdot s=s}
     L_{(s,g)}\Bigr]^G. \tag{1.13} $$
It maps $L_{(s,g)}\otimes L_{(s',g')}$ to zero unless $s'=s$, in which case
it lands in~$L_{(s,gg')}$.  The product is not imposed by hand, but rather is
computed from the functional integral over the sphere with three holes.  (See
\cite{F1,\S6} for the case~$S=\pt$ and~$B=0$.  Beware that the counting
measure~\thetag{1.11} figures in the product, which therefore involves
rational numbers.)
 
In case~$S=G/H$ cocycles $B$ for~$H^3_{G}(G/H;\ZZ)\cong H^3_H(\pt;\ZZ)$ are
equivalent to central extensions~$\Htil$ of~$H$ by~$\TT$.  Then $\scrE$~is
the category of $\zt$-graded representations of~$\Htil$ on which the center
acts by scalar multiplication, and by~\thetag{1.12} $E$~is the complexified
Grothendieck group of equivalence classes.  The description~\thetag{1.13} is
via characters of representations, which are sections of the complex line
bundle over~$H$ associated to~$\Htil\to H$.  The algebra structure is
convolution.  It is a semi-simple algebra which, by classical results of
Frobenius, is diagonalized by the equivalence classes of irreducible
representations.  Notice that before complexification there is no product on
the Grothendieck group~$\Ebar$ of~$\scrE$ (due to the rational numbers in the
counting measure).
 
The theory with $S=G$, where $G$~acts by conjugation, occurs as the
dimensional reduction of the $n=3$~finite TQFT considered in the next
section.  In that case the cocycle for a class in~$H^3_G(G;\ZZ)$ is
transgressed from a cocycle for~$H^4_G(\pt;\ZZ)$, and there is an integral
ring which refines the complex Frobenius algebra of the two-dimensional
theory.
 
The theory with~$G=\{1\}$ is a finite version of the perturbative string
model.  The space~$S$ plays the role of spacetime, and the cocycle~$B$ is the
usual $B$-field of the theory.  The category~$\scrE$ plays the role of the
category of {\it boundary states\/} or {\it D-branes\/} in two-dimensional
conformal field theories.

 \subhead \S{1.8}. The Three-Dimensional Theory
 \endsubhead

The $K$-module~$\scrE=\scrE(\cir)$ attached to the circle in an extended
three-dimensional TQFT has additional structure due to the functional
integral over spheres with disks removed.  We summarize all of this structure
in the assertion that $\scrE$~is a {\it Frobenius algebra
over~$K$\/}.\footnote{It usually goes by other names, such as {\it modular
tensor category\/}.  Of course, in a two-dimensional TQFT functional
integrals over the same spaces lead to the Frobenius algebra structure on the
complex vector space~$E(\cir)$.}  A Frobenius algebra over~$K$ determines a
three-dimensional TQFT~\cite{T1}, much as a Frobenius algebra over~$\CC$
determines a two-dimensional TQFT.  The Grothendieck group~$\Ebar$ of~$\scrE$
is a ring and is known as the {\it Verlinde ring\/} or {\it Verlinde
algebra\/} of the three-dimensional TQFT.  The Frobenius algebra associated
to the dimensional reduction of this theory to two dimensions is the
complexification of the Verlinde algebra.

The most well-known example of a three-dimensional TQFT is the quantum
Chern-Simons theory associated to a compact group~$G$ and a cocycle for a
class in~$H^4_G(\pt;\ZZ)$.  It was introduced by Witten~\cite{W1} (for
connected groups) and subsequently studied by many mathematicians.  The
theory for finite groups~$G$ was studied in~\cite{FQ}, \cite{F1}, and other
places as well.  It is exactly the three-dimensional case of the finite TQFT
described here for~$S=\pt$.  In fact, the generalization to arbitrary~$S$
does not produce anything new: if $S=G/H$ we recover Chern-Simons theory
on~$H$. Consider, then, the theory with $S=\pt$ and $G$~an arbitrary finite
group.  An element of~$H^4_G(\pt;\ZZ)$ determines a Chern-Simons functional
on principal $G$-bundles over closed oriented 3-manifolds, and the choice of
a representing cocycle~$B$ leads to the extended classical action described
earlier.  In particular, there is an equivariant $K$-line bundle over the
groupoid~$\C \cir$ of fields on~$\cir$.  Recall from~\thetag{1.8} that this
is equivalent to a $G$-equivariant $K$-line bundle on~$G$, where $G$~acts on
itself by conjugation.  It may be considered as a cocycle~$\Bhat$ for an
element of~$H^3_G(G;\ZZ)$, obtained from~$B$ by a cochain version of the
transgression
  $$ H^4_G(\pt;\ZZ)\longrightarrow H^3_G(G;\ZZ).  $$
Therefore $\scrE$~is the $K$-module of invariant sections, and its
Grothendieck group, the Verlinde ring, is a twisted equivariant $K$-theory
group: 
  $$ \operatorname{Verlinde}(G,B) \cong  K_G^{0+\Bhat}(G). \tag{1.14} $$
It was computed explicitly in~\cite{F1}, but no connection was made there
with twisted $K$-theory.  The main point of that paper was to use the
extended notions of classical action and quantum invariants to directly
relate Chern-Simons TQFT with quantum groups, at least for finite gauge
groups.  

The connection~\thetag{1.14} with twisted $K$-theory was realized recently,
and it was natural to conjecture that \thetag{1.14}~holds for all compact Lie
groups~$G$ and cocycles~$B$.  There are different precise mathematical
definitions of the left hand side of~\thetag{1.14}.  One possibility is the
free $\ZZ$-module generated by positive energy representations of a central
extension (determined by~$B$) of the loop group of~$G$ endowed with the
fusion product.  The proof of~\thetag{1.14} in this form is ongoing joint
work with Michael Hopkins and Constantin Teleman; see~\cite{F3} for further
motivating remarks and sample computations.  Details will appear elsewhere.

 \newpage
 \head
 Part 2: Anomalies and Pfaffians of Dirac Operators
 \endhead
 \comment
 lasteqno 2@ 19
 \endcomment

The geometric interpretation of the anomaly in the partition function of a
chiral spinor field, or chiral Rarita-Schwinger field,\footnote{The appendix,
joint with Jerry Jenquin, explains some basics about the partition function
of the Rarita-Schwinger field.  These are necessary for the proper
computation of its anomaly.  We remark that {\it nonchiral\/} spinor and
Rarita-Schwinger fields also have (global) anomalies.} was developed in the
mid 1980s.  First, a link with the Atiyah-Singer index theorem was
discovered; see~\cite{AS4} for the first mathematical account.  Witten's
global anomaly formula~\cite{W3}, together with Quillen's work~\cite{Qn} on
the determinant of the $\dbar$-operator on Riemann surfaces, inspired the
development~\cite{BF}, ~\cite{F4} of a geometric structure on the determinant
line bundle of a family of Dirac operators.  See~\cite{C}, \cite{Si2} for
other interpretations and proofs of Witten's formula.  The scope of
mathematical and physical investigations into anomalies during that period
may be gleaned from the conference proceedings~\cite{BW}.
 
In the past 15~years there have been many refinements, extensions, and
variations of these ideas.  We begin here with a general picture of
anomalies, explain why the partition function of a fermionic field is
anomalous, and give a formula for the anomaly in terms of $K$-theory.  The
expression of the full anomaly---including its geometric structure---in {\it
differential $K$-theory\/} is ongoing joint work with M. Hopkins and
I. Singer.

 \subhead \S{2.1}. Actions in Euclidean QFT
 \endsubhead

Consider an $n$-dimensional Euclidean quantum field theory formulated on a
Riemannian $n$-manifold~$X$.  For simplicity we take $X$~to be compact; if it
is not compact we need to impose conditions on the fields at the ends of~$X$.
The space of fields~$\C X$ is in general a groupoid, due to possible gauge
symmetries.  As before, we denote the set of equivalence classes of fields
by~$\Cb X$.  We treat~$\Cb X$ as if it is a smooth manifold.  (In general
there are singularities due to nontrivial automorphisms in the groupoid, and
in a more careful treatment they would be made explicit.)
 
The main character in a lagrangian field theory is the action~$S_X$.  In some
theories only the {\it exponentiated action\/}\footnote{The Euclidean
action~\thetag{2.1} differs from~\thetag{1.9}, which models the action in
Minkowski spacetime, by a factor of~$i$.  See~\cite{F5,Appendix~A} for a
discussion of the Wick rotation which relates them.  We have set~$\hbar=1$,
where $\hbar$~is Planck's constant.}
  $$ e^{-S_X}\:\Cb X \longrightarrow \CC \tag{2.1} $$
is defined; $S_X$~is then not a complex-valued function, but rather a
function with values in the quotient~$\CC/2\pi i\ZZ$.  This suffices as the
{\it quantum partition function\/} is defined by the formal expression
  $$ Z(X) = \int_{\Cb X}\;e^{-S_X}(\varphi )\;\text{``$d\varphi$''}
     . \tag{2.2} $$
More generally, Euclidean quantum field theory consists of integrals of the
exponentiated action times local functionals on the space of fields, called
{\it correlation functions\/} of the local functionals.  For ordinary quantum
field theories $\Cb X$~is infinite-dimensional and much of the mystery of
quantum field theory is hidden in the symbol~``$d\varphi $''.  For our
purposes we treat ``$d\varphi $'' as a measure on~$\Cb X$, so do not discuss
the analytic issues hidden in~\thetag{2.2}.  Instead, we focus on a
variation of~\thetag{2.1} which leads to a geometric issue.
 
One cannot always define the exponentiated action as a complex-valued
function.  Rather, in general it is a product
  $$ e^{-S_X} = \prod\limits_{i} e^{-S_X^{(i)}}  $$
where  
  $$ \text{$e^{-S_X^{(i)}}$ is a section of a geometric line bundle
     $L_X^{(i)}\longrightarrow \Cb X$}.  $$
By ``geometric line bundle'' we understand a complex line bundle endowed with
a metric and compatible unitary connection.  The geometric line
bundles~$L_X^{(i)}$ are part of the data of the classical action.  Thus
  $$ \text{$e^{-S_X}$ is a section of a geometric line bundle
     $L\mstrut _X=\bigotimes\limits_{i}L_X^{(i)}$}.  $$
The integral in~\thetag{2.2} no longer makes sense, as we are attempting to
sum a function which does not take values in a fixed vector space.  To
complete the definition of the quantum theory we need to specify
  $$ \triv X = \text{a trivialization of $L\mstrut _X$}.  $$
The partition function is then 
  $$ Z(X) = \int_{\Cb X}\;\frac{e^{-S_X}}{\triv X}(\varphi
     )\;\text{``$d\varphi$''} ,   $$
which makes sense since the ratio~$e^{-S_X}/\triv X$ is a complex-valued
function on~$\Cb X$.  We require that $\triv X$~be a geometric
trivialization, i.e., it must trivialize both the metric and connection: 
  $$ \aligned
      |\triv X| &=1 \\ 
      \nabla \triv X &=0.\endaligned  $$

This raises existence and uniqueness questions.  The obstruction to finding a
trivialization of~$L\mstrut _X$ is called the {\it anomaly\/} of the
exponentiated action~$e^{-S_X}$.  Notice that the collection of geometric
line bundles over~$X$ forms a groupoid; we denote the set of equivalence
classes by $\Hh^2(X)$.  (The reason for this notation will emerge at the end
of~\S{3.5}.) The set of equivalence classes of {\it topological\/} line bundles
is the integer cohomology group~$H^2(X;\ZZ)$, and there is a natural
forgetful map
  $$ \Hh^2(X)\longrightarrow H^2(X;\ZZ). \tag{2.3} $$
By definition the anomaly is the equivalence class of~$L\mstrut _X$
in~$\Hh^2(X)$.  More specifically, there is an anomaly associated to each
factor in the exponentiated action:
  $$ \Anom(e^{-S_X^{(i)}}) = [L_X^{(i)}]\in \Hh^2(X). \tag{2.4} $$
We say the anomaly cancels if the total anomaly---the sum of
$[L_X^{(i)}]$---is zero.  Sometimes there is a {\it canonical\/}
trivialization of the tensor product of some~$L^{(i)}_X$.\footnote{An example
occurs at the end of the appendix.  A less trivial example is the
Green-Schwarz anomaly cancellation mechanism, in which case the canonical
trivialization follows from the conjectural index theorem in differential
$K$-theory.}
 
The equivalence class of a geometric line bundle~$L$ is determined by the
holonomy around every smooth loop in~$X$.  The holonomy around a contractible
loop is determined from the curvature of~$L$, and the holonomy around a loop
of finite order in~$H_1(X)$ from the curvature and topological class of~$L$.
Therefore, the topological class and/or curvature are often used as
approximations to the anomaly~\thetag{2.4}.  In the physics literature the
curvature is called the {\it local anomaly\/} and the holonomy the {\it
global anomaly\/}.   
 
If the anomaly vanishes, then the set of possible trivializations~$\triv X$
is a torsor for~$H^0(\Cb X;\TT)$, the group of locally constant $\TT$-valued
functions on~$\Cb X$.  (We may divide by the group of constant functions,
since an overall phase does not affect the quantum theory.)  Just as the
exponentiated action~$e^{-S_X}$ is constrained by functoriality and locality,
so too is the trivialization~$\triv X$.  In particular, there are gluing laws
which relate trivializations on different manifolds.  (See~\cite{W2} for an
example in which these constraints play a role.)

 \subhead \S{2.2}. The Partition Function of a Spinor Field
 \endsubhead

In Minkowski spacetime~$\MM^n$ a spinor field is a function $\psi \:\MM^n\to S$
with values in a {\it real\/}\footnote{The reality of~$S$ is essentially the
requirement of CPT~invariance.} spinor representation of the Lorentz
group~$\Spin(1,n-1)$.  The complexification~$S_\CC$ carries a representation
of the complex spin group~$\Spin(n;\CC)$ whose restriction to the Euclidean
spin group~$\Spin(n)$ is in general not real.  So in Euclidean field theory it
makes no sense to impose a reality condition on the spinor fields.  Instead,
the reality condition on spinor fields in Minkowski spacetime ensures that
the Euclidean partition function---the pfaffian of a Dirac operator---makes
sense.\footnote{The fact that the pfaffian of the complexification of a real
matrix equals the pfaffian of the matrix implies that nothing is lost by
complexification.}  The details depend on~$n\pmod8$, so cry out for a unified
treatment using the Bott periodicity of the orthogonal group (or real
Clifford algebras).  We have yet to find a coherent picture along these
lines.
 
Consider a Euclidean field theory on a compact Riemannian $n$-manifold~$X$.
We assume that $X$~is endowed with an orientation and spin structure.  Then
the choice of a real spin representation~$S$ in Minkowski spacetime
corresponds to fixing the rank of a vector bundle~$E$ in the Euclidean
theory.  We consider spinor fields with coefficients in~$E$.  Either $E$~is
fixed to be the trivial bundle of the specified rank, or $E$~is variable in
which case a connection on~$E$ is a field in the theory.  The vector bundle
may carry a real~($\RR$) or quaternionic~($\HH$) structure or it may not be
self-conjugate~($\CC$).  This depends on $n\pmod8$, as indicated in Table~1.
In dimensions~$n=2$ and~$n=6$ there are two vector bundles, corresponding to
the fact that there are two inequivalent irreducible real spinor
representations of the Lorentz groups in those dimensions.  Notice that the
entries in Table~1 also correspond to the reality conditions of the
irreducible complex spinor representation in Lorentz signature~\cite{De}.

 \midinsert
 \bigskip
 \centerline{\eightpoint
$$
\vbox{\offinterlineskip
\hrule
\halign{\vrule# &\quad\hfill #\hfill
  &\quad\vrule# &\quad\hfill #\hfill
  &\quad\vrule# &\quad\vrule# \cr
height6pt &\omit   &&\omit  &\cr
& $n\pmod8$ &&\hfill vector bundle~$E$&\cr
height6pt &\omit   &&\omit &\cr
\noalign{\hrule height 1.5pt depth 0pt}
height3pt &\omit  &&\omit &\cr
&$0$ &&$\CC$  &\cr
height3pt &\omit  &&\omit &\cr
\noalign{\hrule}
height3pt &\omit  &&\omit &\cr
&$1$ &&$\RR$  &\cr
height3pt &\omit  &&\omit &\cr
\noalign{\hrule}
height3pt &\omit  &&\omit &\cr
&$2$ &&$\RR\oplus \RR$  &\cr
height3pt &\omit  &&\omit &\cr
\noalign{\hrule}
height3pt &\omit  &&\omit &\cr
&$3$ &&$\RR$  &\cr
height3pt &\omit  &&\omit &\cr
\noalign{\hrule}
height3pt &\omit  &&\omit &\cr
&$4$ &&$\CC$  &\cr
height3pt &\omit  &&\omit &\cr
\noalign{\hrule}
height3pt &\omit  &&\omit &\cr
&$5$ &&$\HH$  &\cr
height3pt &\omit  &&\omit &\cr
\noalign{\hrule}
height3pt &\omit  &&\omit &\cr
&$6$ &&$\HH\oplus \HH$  &\cr
height3pt &\omit  &&\omit &\cr
\noalign{\hrule}
height3pt &\omit  &&\omit &\cr
&$7$ &&$\HH$  &\cr
height3pt &\omit  &&\omit &\cr
\noalign{\hrule}
}
\hrule}
$$
}
 \nobreak
 \botcaption{Table~1: Reality conditions on~$E$}
 \endcaption
 \bigskip
 \endinsert

For example, the Lorentz group in dimension~$n=3$ is isomorphic
to~$SL(2;\RR)$ and up to isomorphism there is one irreducible real spin
representation~$S$, which has dimension~2.  Thus any real spin representation
of the Lorentz group has the form~$S\otimes _{\RR} E$ for a real vector
space~$E$.  Its complexification $S_{\CC}\otimes _{\CC}E_{\CC}$ is a
representation of the Euclidean spin group, which is isomorphic to~$SU(2)$.
As another example, in dimension~$n=6$ the Lorentz spin group is isomorphic
to~$SL(2;\HH)$, and there are two inequivalent irreducible 2-dimensional
quaternionic spinor representations.  The underlying 8-dimensional real
representations~$S_1,S_2$ are irreducible, and any irreducible real spinor
representation has the form
  $$ S_1\otimes _{\RR}F_1\,\oplus \,S_2\otimes _{\RR}F_2 \tag{2.5} $$
for some real vector spaces~$F_i$.  Now the complexification
of~$S_i$ can be written as~$T_i\otimes W$, where $T_i$~is a 4-dimensional
complex vector space and $W$~is the quaternions, thought of as a
2-dimensional complex vector space.  Note that $T_i$~carries a representation
of the Euclidean spin group, which is isomorphic to~$SU(4)$.  Thus as a
representation of the Euclidean spin group we identify the complexification
of~\thetag{2.5} as 
  $$ T_1\otimes _{\CC}W\otimes _{\CC}(F_1)_{\CC}\,\oplus \,T_2\otimes
     W\otimes _{\CC}(F_2)_{\CC} \cong T_1\otimes _{\CC}E_1 \,\oplus
     \,T_2\otimes E_2  $$
for quaternionic vector spaces~$E_i$ whose complexification is $W\otimes
_{\CC}(F_i)_{\CC}$.

A theory with a spinor field~$\psi $ often involves many other fields~$f$.
We assume that the spinor field enters the action only\footnote{Higher degree
polynomials in~$\psi $ are easily handled as perturbations and do not affect
anomalies.} through a term of the form
  $$ \int_{X}\frac 12 \psi D_f\psi. \tag{2.6} $$
The Dirac operator~$D$ depends on the metric on~$X$ and perhaps also a
connection on~$E$.  The metric and connection may be included among the
fields~$f$, or more generally depend on~$f$.  A mathematical interpretation
of~\thetag{2.6} involves supermanifolds, since $\psi $~is a fermionic field,
but here we only need know that the formal integral of the exponential
of~\thetag{2.6} over~$\psi $ is
  $$ e^{-S_{\text{fermi}}}(f) = \pfaff D_f. \tag{2.7} $$
This is a factor in the exponentiated effective action\footnote{The field
on~$X$ may be written as an infinite dimensional (odd) vector bundle $\Cb
X\to \overline{\Cal{C}_{\text{bos}}(X)}$; the fibers are the fermionic
fields, the base the bosonic fields (denoted~$f$ in the text).  The effective
action is obtained by integration~\thetag{2.2} over the fibers of this map.}
for the fields~$f$.  For fixed~$f$ \thetag{2.7}~makes sense as an element of
a hermitian line
  $$ L_{\text{fermi}}(f) = \Pfaff D_f, \tag{2.8} $$
and as $f$~varies these lines fit together into a smooth geometric line
bundle~$L_{\text{fermi}}$, i.e., a hermitian line bundle with unitary
connection.  The exponentiated effective action~\thetag{2.7} is a section of
~$L_{\text{fermi}}$. 

We recall the formal motivation for~\thetag{2.7} as the integral
of~\thetag{2.6} over~$\psi $.  Replace~$\psi $ by a one-dimensional
real variable~$x$ and the quadratic form~\thetag{2.6} by the quadratic form
$\frac 12\lambda x^2$; then the integral of the exponentiated action over~$x$
is
  $$ \int_{-\infty }^{\infty }dx\,e^{-\frac 12\lambda x^2} \sim \frac
     1{\sqrt\lambda }.  $$
In arbitrary finite dimensions $\lambda $~is replaced by a quadratic form~$Q$
and the right hand side by a multiple of $1/\sqrt{\det Q}$.  (The measure is
used to define the determinant of a quadratic form.)  For a fermionic
variable the answer is the reciprocal of the ordinary integral---it is the
square root of the determinant of~$Q$, i.e., the pfaffian of~$Q$.
 
In {\it odd\/} dimensions the pfaffian line bundle~$L_{\text{fermi}}=\Pfaff
D$ carries a real structure, and the metric and connection are compatible
with it.  Hence over a space of fields~$f$ parametrized by a manifold~$T$ the
line bundle~$\Pfaff D$ is classified up to isomorphism by an element
in~$H^1(T;\zt)$.  In other words, in odd dimensions the curvature vanishes
and the anomaly is topological.  We caution that this topological
interpretation includes the real structure.  The forgetful map which omits
the real structure is the Bockstein homomorphism
  $$ H^1(T;\zt)\longrightarrow H^2(Z;\ZZ),  $$
and the Bockstein of~$\Lferm$ equals the image of~$\Lferm$
under~\thetag{2.3}.

In even dimensions $\Lferm$~ is usually geometrical, so its equivalence class
in~$\Hh^2(T)$ is not determined by passing to a topological cohomology group.

 \subhead \S{2.3}. Construction of~$\Pfaff D$
 \endsubhead

As mentioned earlier there are different constructions of the pfaffian line
bundle depending on the dimension~$n$ modulo~8.  We sketch the cases~$n=1$
and~$n=2$ as illustrative examples.  A more detailed discussion of the odd
dimensional cases appears in~\cite{Si1} and~\cite{MW,\S4.1}.  In
dimensions~$n\equiv 0,4\pmod8$ the pfaffian reduces to an ordinary
determinant.  An extra square root is required for~$n\equiv 2,6\pmod8$.
These even dimensional cases are described in~\cite{F4} and the references
therein.
 
We begin with a finite dimensional analogy.  Let $V$~be a finite dimensional
vector space, real or complex,\footnote{The real case is a model for the
Dirac operator in dimensions~$1,5\pmod8$; the complex case for the Dirac
operator in dimensions~$2,6\pmod8$.  Other finite dimensional models are
relevant for the remaining dimensions.} and
  $$ D\:V\longrightarrow V^* \tag{2.9} $$
a skew-adjoint operator: $D^*=-D$.  This means that under the isomorphism
$\Hom(V,V^*)\cong V^*\otimes V^*$ the operator~$D$ corresponds to $\omega
_D\in {\tsize\bigwedge} ^2V^*$.  Now the pfaffian of~$D$ vanishes if $D$~is
not invertible, and the invertibility of~$D$ requires that $V$~be even
dimensional.  Assuming $\dim V=2r$, we define 
  $$ \pfaff D = \frac{\omega _D^r}{r!}\in \Det V^*, \tag{2.10} $$
where the determinant line of~$V^*$ is~$\Det V^*={\tsize\bigwedge} ^{2r}V^*$.
In this finite dimensional case the pfaffian line~$\Pfaff D$ is~$\Det V^*$,
which is independent of~$D$.
The determinant~$\det D\:\Det V\to\Det V^*$ is the map on highest exterior
powers induced from~\thetag{2.9}.  It is the square of the pfaffian:
  $$ (\pfaff D)^{\otimes 2} = \det D \in (\Det V^*)^{\otimes 2}. 
     $$
If $V$~is real, and $\Det V^*$~is endowed with a metric (i.e. $V$~has a
volume form), then $(\Det V^*)^{\otimes 2}$~is trivial.  In this case the
determinant may be identified with a number, but not the pfaffian.  Without
the metric there is a preferred contractible space of trivializations
of~$(\Det V^*)^{\otimes 2}$, so in a family of operators parametrized by a
manifold~$T$ the line bundle $(\Det V^*)^{\otimes 2}\to T$~is topologically
trivial.  The pfaffian line bundle $\Det V\to T$ has order two.

The pfaffian and pfaffian line may be constructed for skew-adjoint Fredholm
operators~$D$ (see~\cite{Qn}, \cite{S2}).  In this case the pfaffian line
depends on~$D$, not just on the underlying topological vector space.  The
pfaffian line of a Fredholm operator does not carry a natural metric or
connection.
 
A {\it geometric family of Dirac operators\/} parametrized by a manifold~$T$
is given by the following data: a fiber bundle~$\pi \:X\to T$ with
finite-dimensional fibers; a metric on the tangent bundle to the fibers
of~$\pi $; a horizontal distribution on~$\pi $, i.e., a distribution in~$TX$
complementary to the relative tangent bundle~$T(X/T)\subset TX$; an
orientation and spin\footnote{To construct complex Dirac operators a
$\text{spin}^c$ structure suffices.  In most cases we require a spin
structure to construct the pfaffian, though variations are possible in some
situations.} structure on~$T(X/T)$; and a vector bundle $E\to X$ with
connection.  In addition we assume that the fibers of~$\pi $ are closed
(compact without boundary).  Then the Dirac operator has discrete spectrum
and the eigenvalues grow in absolute value at a rate depending on the
dimension.  This leads to a well-defined heat kernel and ultimately to
geometric invariants.

Consider first the case where the fibers of~$\pi $ have dimension~$n=1$.  The
Dirac operator~$D$ is a real skew-adjoint Fredholm operator, so there is a
general construction of the pfaffian line bundle in a family.  However, we
need a {\it geometric\/} line bundle, so must examine~$D$ in more detail.
Relative to appropriate trivializations we may identify~$D$ with the operator
~$\frac{d}{dx} + A$, where $A$~is a constant skew-adjoint matrix and $x$~a
local coordinate.  The spectrum is symmetric about the origin, and if the
operator is invertible\footnote{There is a mod~2 index, which is locally
constant on~$T$, which if nonzero guarantees the existence of a zero
eigenvalue.  In that case the pfaffian is defined to be identically zero.
This is analogous to the requirement in finite dimensions that the dimension
of the underlying vector space of a skew-adjoint operator be even.} the
absolute value of the pfaffian, which is formally
  $$ |\pfaff D| = \prod\limits_{\overset{\lambda \in \spec D}\to{ |\lambda
     |>0}} |\lambda|, \tag{2.11} $$
is defined using the heat kernel, or equivalently by analytic continuation of
an appropriate $\zeta $-function~\cite{RS}.  But the sign of the pfaffian is
not defined.  As we move in the parameter space we encounter zeros in the
spectrum, and as pairs of eigenvalues pass through zero the sign becomes
ill-defined.  More precisely, suppose we look along a path in~$T$ which has
an isolated zero, say at $s=0$ for a parameter~$s$ along the path.  Then to
leading order $|\pfaff D_s| \sim cs^k$, where $\dim\ker D(0)=2k$.  A {\it
smooth\/} choice of a real-valued function~$\pfaff D$ must change sign
at~$s=0$ if $k$~is odd and must not change sign at~$s=0$ if $k$~is even.
Thus there is a smooth function~$\pfaff D$ with absolute value~\thetag{2.11}
if and only if for each loop (with appropriate transversality) the total
number of pairs of eigenvalues which cross zero is even.  The function which
counts such zeros is an element of~$H^1(T;\zt)$.
 
A construction which yields a smooth line bundle $\Pfaff D\to T$ and a
section~$\pfaff D$ proceeds as follows.  For each nonnegative real number~$a$
let $U_a\subset T$ be the subset of parameter values~$t\in T$ for which $\pm
a\notin \spec D_t$.  Over~$U_a$ there is a finite dimensional real vector
bundle $V_a\to U_a$ whose fiber at~$t\in T$ is the sum of eigenspaces
of~$D_t$ for eigenvalues of absolute value less than~$a$.  Then $D$~restricts
to a family of skew-adjoint operators on~$V_a$, so the finite dimensional
construction yields a real line bundle ~$L_a\to U_a$ with a section.  There
is a natural isomorphism~$L_a\cong L_b$ over~$U_a\cap U_b$, which is the
pfaffian of~$D$ restricted to the sum of eigenspaces for eigenvalues of
absolute value between~$a$ and~$b$, and the sections of~$L_a$ and~$L_b$ agree
under this isomorphisms.  These isomorphisms patch $L_a\to U_a$ into a smooth
line bundle $\Pfaff D\to T$ and the sections into a smooth section~$\pfaff
D$.  One uses the zeta function regularization of~\thetag{2.11} to construct
a metric on~$\Pfaff D$; see~\cite{Q}, \cite{BF}, \cite{F4}.  A real line
bundle with a metric has a unique compatible connection.  The topological
equivalence class of~$\Pfaff D$ in~$H^1(T;\zt)$ agrees with that in the
previous paragraph.
 
Now consider the case~$n=2$, so a family of compact oriented spin Riemannian
surfaces parametrized by~$T$.  Such surfaces have a complex structure and
also a square root~$K^{1/2}$ of the canonical bundle.  The (chiral) Dirac
operator~$D$ is identified with the $\dbar$~operator coupled
to~$K^{1/2}\otimes E$:
  $$ \text{\hskip 1in$D = \dbar(K^{1/2}\otimes E)\:$\lower.3 truein\vtop{$
      \CD \Omega ^{0,0}(K^{1/2}\otimes E)@>>> \Omega ^{0,1}(K^{1/2}\otimes
     E)\\
      @| @| \\
      \Omega ^{1/2,0}(E) @>>> \Omega ^{1/2,1}(E)\endCD$}}  $$
The domain and codomain are naturally dual---the duality pairing is the
integral of the product---and by Stokes' theorem the Dirac operator is
skew-adjoint.  This family of complex skew-adjoint Fredholm operators
determines a complex Pfaffian line bundle over~$T$, but without a metric and
connection.  For a geometric construction we use a covering~$\bigl\{U_a
\bigr\}t_{a>0}$ of~$T$ as above, where now
  $$ U_a = \{t\in T: \text{$a$ is not in $\spec D^*_tD_t$}\}.  $$
As before, on~$U_a$ each Dirac operator~$D$ restricts to a skew-adjoint
operator on the sum of the eigenspaces of~$D^*D$ with eigenvalue less
than~$a$; the codomain is the corresponding sum of eigenspaces of~$D^*D$.
These finite dimensional vector spaces are dual if the numerical index of the
Dirac operator vanishes; if not, then the pfaffian is identically zero.
There is a patching construction of~$\Pfaff D$ with metric and connection.
As before we use zeta functions to regularize infinite products and infinite
sums.  Since $\Pfaff D$~is complex, the connection does not follow
automatically from the metric, and more linear analysis is needed~\cite{BF}, \cite{F4}.

 \subhead \S{2.4}. Computation of the Topological Anomaly
 \endsubhead

Topological invariants of families of Dirac operators were thoroughly
investigated by Atiyah and Singer~\cite{AS1}.  The papers of particular
relevance are~\cite{AS2} and~\cite{AS3}.  In even dimensions the
Atiyah-Singer index theorem gives a formula for the topological equivalence
class of the line bundle~$L_{\text{fermi}}$ in~$H^2(T;\ZZ)$.  In odd
dimensions it gives a formula for the topological equivalence class
in~$H^1(T;\zt)$ of~$L_{\text{fermi}}$ with its real structure.  Both are
formulas in $K$-theory, or perhaps in the more refined $KO$- and
$KSp$-theories.\footnote{In low dimensions the $K$-theory formulas below may
be expressed in cohomological terms, starting with a characteristic class of
the vector bundle~$E$.  For example, if~$n=1$ and we restrict to oriented
bundles~$E$, the $K$-theory formula reduces to the pushforward of the second
Stiefel-Whitney class of~$E$ in cohomology~\cite{FW,\S5}.  (There does not
appear to be a cohomology formula if $E$~is not oriented.)  For ~$n=2$ there
is a theorem along these lines in~\cite{F4,\S5} if $E$~is a virtual bundle of
rank zero and is oriented.}  The formula depends on~$n\pmod 8$, where $n$~is
the dimension of the fibers of~$\pi \:X\to T$, though the story for~$i\pmod8$
is quite similar to the story for~$i+4\pmod8$.  We simply summarize the
formulas here.  In each case the formula is expressed in terms of the
equivalence class~$[E]$ of the vector bundle $E\to X$ in the appropriate
$K$-theory group and the pushforward~$\pi _!$ in the appropriate $K$-theory
constructed from the orientation and spin structure on the fibers of~$\pi $.
Recall that~$E$ is real, complex, or quaternionic as indicated in Table~1.

 \subsubhead $\bold{n\cong 1\pmod8}$ \endsubsubhead In this case $E$~is real,
so $[E]\in KO^0(X)$.  Suppose $n=8r+1$.  The equivalence class of the
pfaffian line bundle with its real structure is the image of~$[E]$ under the
sequence of maps
  $$ KO^0(X) @>{\pi _!}>> KO^{-(8r+1)}(T) @>>> H^1(T;\zt). \tag{2.12} $$%KEEPTAG 
One model for an element of~$KO^{-1}(T)$ is a homotopy class of maps $T\to
O(\infty )$, and there is a universal class in~$H^1\bigl(O(\infty );\zt
\bigr)$ which defines the second map in~\thetag{2.12}.

 \subsubhead $\bold{n\cong 2\pmod8}$ \endsubsubhead In this case there are
two real bundles~$E_1,E_2$ over~$X$, corresponding to the two types of spinor
fields in Minkowski spacetime.  Then $[L_{\text{fermi}}]$~is the image
of~$[E_1] - [E_2]$ under the sequence of maps
  $$ KO^0(X) @>{\pi _!}>> KO^{-(8r+2)}(T) @>>> H^2(T;\ZZ). \tag{2.13} $$%KEEPTAG 
This last map is properly called the pfaffian line bundle; it fits into a
commutative diagram 
  $$ \CD 
      KO^{-2}(T) @>\Pfaff>> H^2(T;\ZZ)\\ 
      @V\otimes \CC VV @VV\times 2V \\ 
      K^{-2}(T) @>\Det>> H^2(T;\ZZ)\endCD  $$

 \subsubhead $\bold{n\cong 3\pmod8}$ \endsubsubhead $[L_{\text{fermi}}]$~is the
image of~$[E]$ under the sequence of maps
  $$ KO^0(X) @>{\pi _!}>> KO^{-(8r+3)}(T) @>>> H^1(T;\ZZ) @>>>
     H^1(T;\zt). \tag{2.14} $$%KEEPTAG 
The second map may be understood analytically as the exponential of $\pi
i/2$~times the $\eta $-invariant; of course, there is a topological
interpretation as well.  The analytic interpretation enters into the
definition of~$\pfaff D$ (see~\cite{Si1}).  The last map is simply reduction
modulo two.

 \subsubhead $\bold{n\cong 4\pmod8}$ \endsubsubhead $[L_{\text{fermi}}]$~is
the image of~$[E]$ under the sequence of maps
  $$ K^0(X) @>{\pi _!}>> K^{-(8r+4)}(T) @>>> H^2(T;\ZZ). \tag{2.15} $$%KEEPTAG 
In this case $\pfaff D$~is simply the determinant of the Dirac operator
coupled to~$E$, and \thetag{2.15}~computes this determinant.

 \subsubhead $\bold{n\cong 5\pmod8}$ \endsubsubhead The next three cases may
be related to the first three using Bott periodicity $KSP^{i} \cong
KO^{i+4}$.  Thus $[L_{\text{fermi}}]$~is the image of~$[E]$ under the
sequence of maps
  $$ KSp^0(X) @>{\pi _!}>> KSp^{-(8r+5)}(T) @>>> H^1(T;\zt). 
     \tag{2.16} $$%KEEPTAG 

 \subsubhead $\bold{n\cong 6\pmod8}$ \endsubsubhead $[L_{\text{fermi}}]$~is
the image of~$[E_1] - [E_2]$ under the sequence of maps
  $$ KSp^0(X) @>{\pi _!}>> KSp^{-(8r+6)}(T) @>>> H^2(T;\ZZ).  
     \tag{2.17} $$%KEEPTAG 

 \subsubhead $\bold{n\cong 7\pmod8}$ \endsubsubhead $[L_{\text{fermi}}]$~is the
image of~$[E]$ under the sequence of maps
  $$ KSp^0(X) @>{\pi _!}>> KSp^{-(8r+7)}(T) @>>> H^1(T;\ZZ) @>>>
     H^1(T;\zt). \tag{2.18} $$%KEEPTAG 

 \subsubhead $\bold{n\cong 8\pmod8}$ \endsubsubhead $[L_{\text{fermi}}]$~is
the image of~$[E]$ under the sequence of maps
  $$ K^0(X) @>{\pi _!}>> K^{-(8r+8)}(T) @>>> H^2(T;\ZZ). \tag{2.19} $$%KEEPTAG 

 \subhead \S{2.5}. Computation of the Geometric Anomaly
 \endsubhead

As mentioned several times, in odd dimensions the equivalence class of the
anomaly~$L_{\text{fermi}}$ is topological, so the Atiyah-Singer theorem
suffices.  But in even dimensions there is geometric information, and one
needs a geometric refinement of the Atiyah-Singer theorem to compute the class
of~$L_{\text{fermi}}$ in~$\Hh^2(T)$.  The class is determined by the holonomy
of~$L_{\text{fermi}}$ about all loops in~$T$.  Witten's global anomaly
formula~\cite{W3} expresses the holonomy in terms of exponentiated $\eta
$-invariants.  See~\cite{BF}, ~\cite{F4} for the interpretation (and proof)
in terms of determinant and pfaffian line bundles, and~\cite{Si2}, ~\cite{C}
for other interpretations and proofs.  The curvature of~$L_{\text{fermi}}$,
or local anomaly, has a local formula in terms of the curvature of the metric
on $X\to T$ and the curvature of~$E\to T$.
 
Just as there is a refinement of $H^2(T;\ZZ)$ to~$\Hh^2(T)$, so too there are
refinements of topological $K$-theory groups~$K^i(T)$ to {\it differential
$K$-theory groups\/}~$\Kh^i(T)$.  (There are similar refinements for real and
quaternionic $K$-theory, indeed for any generalized cohomology theory.)  An
ongoing project with Michael Hopkins and Isadore Singer is designed to
interpret invariants of geometric families of Dirac operators as cocycles for
differential $K$-theory.  From this point of view a vector bundle with
connection is a cocycle for a differential $K$-theory class, and the
sequences of maps in the previous section have refinements in the
differential version.  The theory to be developed will imply that the
geometric anomaly (in even dimensions) is computed by the refined sequences
of maps.

 \newpage
 \head
 Part 3: Abelian Gauge Fields
 \endhead
 \comment
 lasteqno 3@ 35
 \endcomment

In \S{1.2} we met a finite version of gauge fields, namely principal bundles
with finite structure group.  In this section we discuss principal bundles
whose structure group is abelian.  Furthermore, they are endowed with a
connection, which is usually called the {\it gauge field\/} or {\it gauge
potential\/}.  In classical electromagnetism the gauge group is the
translation group~$\RR$, and the connection is determined up to isomorphism
by its curvature.  As we recall this curvature, or {\it field strength\/},
encodes the electric and magnetic fields.  The entire classical theory may be
expressed in terms of the field strength; the gauge field plays only a formal
role.  On the other hand, in quantum theories the gauge field is essential.
Furthermore, the gauge group is~$\RR/q\ZZ$ for an appropriate~$q\in \RR$.
This compactification of the gauge group from~$\RR$ to~$\RR/q\ZZ$ is
equivalent to the quantization of electric charge.  Dirac's argument for
charge quantization depends on the existence of nonzero magnetic current,
which again is a feature of the quantum theory not found in classical
electromagnetism.  This story can be told with some variations in arbitrary
dimensions and with field strengths which are differential forms of arbitrary
degree.  One particularly illuminating case, in which the geometry is more
apparent, is when the field strength has degree one.  Then the ``gauge
field'' is a (twisted) map to the circle.  In this case we illustrate
concretely an anomaly in the electric coupling, which occurs when there is
simultaneous magnetic and electric current.

In higher degrees the gauge field is a differential geometric object---a
generalized differential cocycle---whose precise nature depends on the
quantization law for charges.  

In superstring theories it turns out that the appropriate quantization law
for electric and magnetic Ramond-Ramond charges, as well as fluxes, is some
form of $K$-theory.  The electric coupling anomaly in these cases is
expressed in terms of differential $K$-theory.  As explained in~\S{2.5}
fermion anomalies are also (conjecturally) expressed in differential
$K$-theory, so can potentially cancel an electric coupling anomaly.  Such a
cancellation is termed the {\it Green-Schwarz mechanism\/}.
 
Different expositions of this material, and further information about the
superstring examples, is contained in~\cite{FH} and~\cite{F5}.  Here we
emphasize the elementary pictures which motivate that work.

 \subhead \S{3.1}. Maxwell's Equations
 \endsubhead

We work on a four-dimensional spacetime of the form $M^4 = \EE^1\times N^3$,
where $(N^3,g_N)$~is a Riemannian manifold.  We endow~$M$ with the Lorentz
metric $dt^2 - g_N$, where $t$~is a (time) coordinate on~$\EE^1$ and the
speed of light is set to unity.  Minkowski spacetime is the case~$N=\EE^3$.
 
Classical electromagnetism involves four fields: 
  $$ \alignedat 2
      E&\in \Omega ^1(N) &&\qquad \qquad \text{electric field} \\
      B&\in \Omega ^2(N) &&\qquad \qquad \text{magnetic field} \\
      \rho _E&\in \Omega ^3_c(N) &&\qquad \qquad \text{electric charge
     density} \\
      J_E&\in \Omega ^2_c(N) &&\qquad \qquad \text{electric
     current}\endaligned  $$
Here $\Omega _c$~denotes differential forms of compact support.  Traditional
texts identify~$E,B,J_E$ with vector fields and $\rho _E$~with a
function.\footnote{This assumes that $M$~is oriented.  For simplicity we
assume that spacetime~$M$---and later in the Euclidean version the Riemannian
``spacetimes''~$X$---are oriented.  If not, then $\rho _E, J_E$~are forms
twisted by the orientation bundle.  The vector field which corresponds to~$B$
is also twisted.}  But the differential form language is more convenient and
leads to a better geometric picture.  The classical Maxwell equations are
  $$ \alignedat2
      dB&=0&\qquad \qquad \frac{\partial B}{\partial t} + dE &=0 \\
      d*\mstrut _NE &= \rho _E &\qquad \qquad *\mstrut _N\frac{\partial
     E}{\partial t} - d*\mstrut _NB &= J_E\endaligned \tag{3.1} $$
We reformulate these equations using differential forms on~$M$ with its
Lorentz metric and corresponding Hodge $*$~operator as follows.  Set 
  $$ \alignedat2
      F &= B - dt\wedge E &&\qquad \in \Omega ^2(M) \\
      j_E &= \rho _E + dt\wedge J_E &&\qquad \in \Omega ^3(M).\endaligned
      $$
The {\it electric current\/}~$j_E$ has compact spatial support.  Maxwell's
equations~\thetag{3.1} are equivalent to the pair of equations 
  $$ \aligned
      dF &=0 \\
      d*F &= j_E.\endaligned \tag{3.2} $$
As a consequence of the second equation we have 
  $$ dj_E=0. \tag{3.3} $$
Equation~\thetag{3.3} leads to conservation laws through the use of Stokes'
theorem.  
 
There is a global condition~\thetag{3.8} which needs to be added
to~\thetag{3.2} in the classical theory; we discuss it below.

        \remark{\protag{3.4} {Remark}}
 There is an asymmetry in~\thetag{3.2} which would be corrected if we
postulate a {\it magnetic current\/} $j_B\in \Omega ^3(M)$ of compact spatial
support and replace the first Maxwell equation with 
  $$ dF = j_B. \tag{3.5} $$
In the classical theory there is no magnetic current, but the quantum theory
allows for it and, as we explain below, leads to the quantization of both
electric and magnetic charge.  If~\thetag{3.5} were admitted in the classical
theory, it would obey the symmetry of {\it electomagnetic duality\/}: 
  $$ \aligned
      F \longleftrightarrow *F \\
      j_B \longleftrightarrow j_E.\endaligned  $$
There is a quantum version of this symmetry, but for the classical
equations~\thetag{3.2} it only holds in a vacuum ($j_E=0$). 
        \endremark

Let $\iota _t\:N\hookrightarrow M$ be the inclusion at time~$t$.  Then
\thetag{3.3}~implies that the de Rham cohomology class 
  $$ \bQ_E = [\iota _t^*j_E]\quad \in H^3_c(N;\RR)  $$
is independent of~$t$.  It is called the total {\it electric charge\/}.  If
$N$~is connected, then $H^3_c(N;\RR)\cong \RR$, and the total charge is a
real number.  Notice that the second Maxwell equation~\thetag{3.2} implies
  $$ \bQ_E \in \ker\bigl(H^3_c(N;\RR)\longrightarrow H^3(N;\RR)
     \bigr). \tag{3.6} $$
In particular, it vanishes if $N$~is compact. 
 
A typical source for electric current is a collection of charged particles.
They may be fixed background objects, too heavy to be affected by the
electromagnetic field~$F$, or may be dynamical objects whose equations of
motion are coupled to Maxwell's equations.  Let the worldlines of the
particles be a submanifold $i\:W^1\hookrightarrow M$.  We don't assume $W$~is
connected, so allow for several particles, but do assume that for each~$t$
the intersection $W^1\cap \bigl(\{t\}\times N \bigr)$~is compact.
Physically, we should also assume that $W$~is timelike, though this is not
essential for the issues at hand.  The electric charges are specified by a
locally constant function $q_E\:W\to \RR$, that is $q_E\in \Omega ^0(W)$
with~$dq_E=0$.  The induced electric current may be written\footnote{Various
``twistings'', due to orientations or put in by hand, may be present in the
fields; see the end of~\cite{F5,\S2}.}
  $$ j_E = i_*q_E\qquad \in \Omega ^3(M). \tag{3.7} $$
The most straightforward interpretation of~\thetag{3.7} is as a
distributional differential form---a de Rham current---but we prefer to use a
smooth representative to avoid illegal products of distributions.  Of course,
this smoothing involves a choice.  In any case the electric charge~$\bQ_E\in
H^3_c(N;\RR)$ is independent of the choice.  Notice that the current~$j_E$,
which is a differential-geometric quantity, encodes the positions,
velocities, and charges of the individual particles, whereas the
charge~$\bQ_E$, which is a topological quantity, only encodes the total
charge.

We can generalize this classical picture easily in two directions.  First, we
may replace~$N^3$ with a Riemannian manifold~$N^{n-1}$ of arbitrary
dimension.  Secondly, we may replace the electromagnetic field~$F$ with a
differential form of arbitrary degree~$d$.  There are corresponding changes
in the degrees of~$j_E$ (and of~$j_B$ in the quantum theory).  We can go
further and consider a collection of forms~$F=(F_1,\dots ,F_k)$ of
multidegree~$\bold{d}=(d_1,\dots ,d_k)$.  But for the moment we continue with
a single field of degree~$d=2$ in $n=4$~dimensions.

 \subhead \S{3.2}. The Action Principle for Electromagnetism
 \endsubhead

We treat Maxwell's equations~\thetag{3.2} asymmetrically to write an action
principle.  First, we need to add to the first Maxwell equation the condition 
  $$ [F]=0\qquad \in H^2(M;\RR). \tag{3.8} $$
Of course, for $N=\EE^3$ this holds automatically, but for example if
$N=\pnct$ it is a nontrivial condition.  It ensures that all periods of~$F$
around 2-cycles in~$N$ vanish, which physically means that there are no
magnetic charges.  As we already discussed magnetic charges are excluded in
the classical theory.  It follows that there exist 1-forms~$A\in \Omega
^1(M)$ such that
  $$ F=dA. \tag{3.9} $$
Furthermore, the {\it gauge field\/} or {\it gauge potential\/}~$A$ is
determined up to addition of closed forms.  The space of classical fields is
then the quotient space
  $$ \scrF_{\text{classical}} = \frac{\Omega ^1(M)}{\clf ^1(M)}, \tag{3.10} $$
where $\clf$~denotes closed differential forms.  The differential~$d$ maps it
isomorphically onto the space of exact 2-forms, in other words to the space
of electromagnetic fields~$F$.

        \remark{\protag{3.11} {Remark}}
 We can reformulate the gauge field in the language of principal bundles and
connections.  Namely, take~$A$ to be a connection on a principal bundle
over~$M$ whose structure group is the real numbers~$\RR$ (under addition).
The space of these connections up to equivalence is an affine space based
on~$\Omega ^1(M)/d\Omega ^0(M)$.  The quotient by equivalence classes of flat
connections is an affine space based on~\thetag{3.10}.  This is the correct
space for classical electromagnetism; no ``classical experiment''
distinguishes between fields which differ by a flat connection.  In the
quantum theory, however, there are such experiments (the Aharonov-Bohm
effect).  
        \endremark

Equation~\thetag{3.9} ensures that the first Maxwell equation ~\thetag{3.2}
is satisfied.  The second Maxwell equation is the Euler-Lagrange equation of
the action
  $$ S(A) = \int_{M}\;-\frac 12dA\wedge *dA + A\wedge j_E,\qquad A\in \Omega
     ^1(M). \tag{3.12} $$
It is {\it not\/} the case that $S$~is well-defined on the
quotient~\thetag{3.10}.  However, the integrand (lagrangian) is well-defined
up to an exact term.  Namely, it follows from~\thetag{3.6} that $j_E=dG$ for
a global 2-form~$G$.  (We can take $G=*F$ for any solution~$F$ to the
classical equations, but we do not need $G$~to solve any equations for this
argument, which is topological in nature.)  Then \thetag{3.12}~may be
rewritten as
  $$ S(A) = \int_{M}\;-\frac 12dA\wedge *dA + dA\wedge G - d(A\wedge
     G). \tag{3.13} $$
This shows that $S$~depends only on~$dA$, that is, the image of~$A$ in the
quotient~\thetag{3.10}, up to an exact term.  That is all we require of an
action in classical physics, as the Euler-Lagrange equations are unaffected
by exact terms.\footnote{The Euler-Lagrange equations are derived by
considering {\it compactly supported\/} variations~$\dot A$ of~$A$, and for
these $\int_{M}d(\dot A\wedge G)=0$.}  Now a straightforward computation
shows that the Euler-Lagrange equation of~\thetag{3.13} is the second Maxwell
equation~\thetag{3.2}.

 \subhead \S{3.3}. Dirac Charge Quantization
 \endsubhead

To write a quantum mechanical theory which incorporates
electromagnetism---for example, the nonrelativistic Schr\"odinger equation
for a charged particle moving in a background electromagnetic field---the
gauge potential~$A$, and not just the electromagnetic field~$F=dA$, appears.
This assertion has an experimental basis, due to Aharanov and Bohm.
Furthermore, it is an empirical fact that nobody has written a quantum theory
in terms of $F$~alone; see~\cite{Fyn,\S II-15-5} for a discussion.  Accepting
the necessity of the gauge potential, the quantization of charge is based on:
(i)\ the existence of a system in which the magnetic current~$j_B$ and
electric current~$j_E$ are both nonzero, and (ii)\ the particular coupling
of~$A$ to the electric current in the quantum theory.
 
We continue with standard electromagnetism on~$M^4 = \EE^1\times N^3$, and
assume that $N=\EE^3$ is standard Euclidean space.  Suppose there is a static
{\it magnetic monopole\/} of magnetic charge~$q_B$ at the origin of space.
We represent it as a magnetic current 
  $$ j_B = q_B\cdot \delta  $$
localized at the origin.  Here $\delta $~is the distributional 3-form
on~$M^4$ dual to the 1-manifold $\EE^1\times \{0\}\subset \EE^1\times
\EE^3$.  The first Maxwell equation, modified as in~\thetag{3.5} to
incorporate magnetic current, is then 
  $$ dF = q_B\cdot \delta . \tag{3.14} $$
We can no longer write $F=dA$ globally on~$M$; the right-hand side
of~\thetag{3.14} is a local obstruction to writing~$F=dA$.  Although it is
eventually important to us to define the gauge field~$A$ on all of spacetime,
we first try to define it on subsets.  For example, on $\EE^1\times
\bigl(\pnct \bigr)$ there is no local obstruction, but now there is a global
obstruction
  $$ \int_{S}F = q_B, \tag{3.15} $$
where, say, $S$~is the unit 2-sphere in~$\{t\}\times \bigl(\pnct \bigr)$ for
any~$t$.  Dirac eliminates this global obstruction by introducing a ray~$R$
emanating from the origin~$0\in \EE^3$, a so-called ``Dirac string''.  (A
physical model is a semi-infinite solenoid.)  Then $A$~can be defined
on~$\EE^1\times (\EE^3\setminus R)$ and Dirac's argument is based on the
premise that the string be invisible.  In essence, it is based on the global
obstruction~\thetag{3.15} due to $H^2\bigl(\pnct \bigr)\not= 0$.  However one
phrases the argument, there is a tension between the necessity of a local
gauge potential~$A$ in quantum theory and the global topological obstruction
to its existence as a 1-form.
 
Differential geometry provides a well-known relief of this tension.  We
emphasize, however, that there is a choice\footnote{The choice is based
partly on the form of the electric coupling in the theory, and partly on
other physical considerations (anomaly cancellation, for example).
See~\S{3.7} for a discussion about this choice for Ramond-Ramond charges in
superstring theory.}, and in other theories different differential-geometric
constructions may enter at this point.  Namely, we take $F$~to be the
curvature of a connection~$A$ on a principal $\RR/q_B\ZZ$-bundle $\pi \:P\to
\EE^1\times \bigl(\pnct \bigr)$.  Thus, $A$~is a right-invariant form on~$P$
which satisfies $dA=\pi ^*F$.  The global condition~\thetag{3.15} implies
that $P$~represents the generator of~$H^2\bigl(\EE^1\times (\pnct) \bigr)$.
The classical space of fields~\thetag{3.10} is now replaced by the quantum
groupoid of fields
  $$ \scrF_{\text{quantum}} = \text{category of principal
     $\RR/q_B\ZZ$-bundles with connection}.  $$
Morphisms are connection-preserving bundle isomorphisms.  The space of
equivalence classes~$\overline{\scrF_{\text{quantum}}}$ fits into an exact
sequence
  $$ 0 \longrightarrow H^1(\;\cdot\; ;\RR/q_B\ZZ) \longrightarrow
     \overline{\scrF_{\text{quantum}}} \longrightarrow
     \scrF_{\text{classical}} \longrightarrow 0.  $$
On the spacetime $M=\EE^1\times \bigl(\pnct \bigr)$ we have
$\overline{\scrF_{\text{quantum}}}= \scrF_{\text{classical}}$, since the
first cohomology vanishes, but in general
$\overline{\scrF_{\text{quantum}}}$~is an extension
of~$\scrF_{\text{classical}}$ by the equivalence classes of flat
connections.  This fits the fact that flat connections can be detected in
quantum mechanics, but not in classical physics. 
 
The Dirac quantization law follows from the requirement that the
exponentiated {\it electric coupling\/}
  $$ \exp\left( \frac{i}{\hbar}\int_{M}A\wedge j_E \right) \tag{3.16} $$
be well-defined.  Here $\hbar$~is Planck's constant, which appears in any
quantum theory.  This expression is a factor in the exponentiated action
which enters the functional integral formulation of quantum mechanics and
quantum field theory.  The same expression appears in the exponentiated
Euclidean action, except that $M=\EE^1\times N$ is replaced by an arbitrary
Riemannian 4-manifold~$X$.  In the Euclidean framework, to which we now turn,
$A$~is a connection in an $\RR/q_B\ZZ$-bundle $P\to X$.  For now we do not
consider general electric currents~$j_E$, but rather the current associated
to a ``Wilson loop'' of electric charge~$q_E$.  That is, suppose as
in~\thetag{3.7} that $j_E$~is $q_E$~times the Poincar\'e dual to an oriented
1-manifold $W^1\subset X$, which we now assume is closed.  Then
\thetag{3.16}~reduces to
  $$ \exp\left( \frac{i}{\hbar}\;q_E\!\int_{W}A \right).  \tag{3.17} $$
We interpret the integral as the holonomy of~$A$ around~$W$, which takes
values in~$\RR/q_B\ZZ$.  Thus \thetag{3.17}~is well-defined if 
  $$ \frac{q_Eq_B}{\hbar} \in 2\pi\ZZ. \tag{3.18} $$
Equation~\thetag{3.18} is Dirac's quantization law.   
 
The same conclusion may be derived in the nonrelativistic quantum mechanics
of a particle of charge~$q_E$ on~$N=\pnct$.  The wave function of the
charged particle is a section of the line bundle associated to the
representation 
  $$ \aligned
      \RR/q_B\ZZ &\longrightarrow \TT \\
      x &\longmapsto \exp\left( \frac{i}{\hbar}q_Ex \right), \endaligned
      $$
which is well-defined only if~\thetag{3.18} is satisfied. 
 
In either argument we see directly that the compactness of the gauge group
immediately implies the quantization of charge.

Summarizing, the argument for charge quantization appears in two stages.
First, a nonzero magnetic current~$j_B$ alters the global nature of the gauge
potential~$A$.  Second, the electric coupling~\thetag{3.17} forces a
quantization law on the electric current~$j_E$, and so on the electric
charge.  Observe that the global nature of~$A$ depends on the magnetic
current~$j_B$, a result of which is that the quantization law involves the
product of magnetic and electric charges.

 \subhead \S{3.4}. The Electric Coupling Anomaly
 \endsubhead

We begin to move away from the Maxwell theory toward more general abelian
gauge fields.  From now on we work in the Euclidean framework, so over an
oriented Riemannian manifold~$X$.  (In fact, the Riemannian metric is
irrelevant in this discussion, but still it is better viewed in the framework
of Euclidean quantum field theory.)  We assume $X$~is compact to avoid
convergence issues.  Let $X$~ have arbitrary dimension~$n\ge 2$.  The
geometry of the anomaly is clearest for a single field strength~$F$ of
degree~$d=1$.  Classically, then, $F$~is an exact 1-form on~$X$ and the
classical gauge field
  $$ A_{\text{classical}} \in \frac{\Omega ^0(X)}{\clf^0(X)}  $$
is a function up to locally constant functions.   
 
In the quantum theory we admit a nonzero magnetic current~$j_B$, which is a
closed 2-form on~$X$.  The case $\dim X=2$ is easiest to visualize.  Then let
$i_B\:W_B\hookrightarrow X$ to be the inclusion of a finite set of oriented
points, and $q_B\:W_B\to \RR$ the magnetic charges of these points.  The
induced magnetic current~$j_B = (i_B)_*(q_B)$ is $q_B$~times a smooth
Poincar\'e dual form to~$W_B$.  Normalize\footnote{There is a nontrivial
restriction on the values of~$q_B$, namely that they be integer multiples of
a nonzero real number.  If not, we would not be able to define the gauge
potential, at least in this framework.}  the magnetic charges so that they
lie in~$2\pi \ZZ$:
  $$ q_B\:W_B\longrightarrow 2\pi \ZZ; \tag{3.19} $$
then $j_B/2\pi $~has integral periods.  Represent the magnetic current~$j_B$
geometrically\footnote{This is a differential-geometric version of the
construction of a holomorphic line bundle on a complex curve from a divisor.}
by a principal $\RR/\tpz$-bundle $\scrJ_B\to X$ with connection whose curvature
is~$j_B$.  This is a refinement of the differential form current~$j_B$, as
$j_B$~does {\it not\/} determine~$\scrJ_B$ up to isomorphism
if~$H^1(X;\RR/\tpz)\not= 0$.  

In the previous section we defined the gauge potential~$A$ outside the
support of~$j_B$, and here we begin similarly.  Set $X^0=X\setminus \supp
j_B$.  Now the quantization condition~\thetag{3.19} leads to the fact that
$[j_B/2\pi]\in H^2(X;\RR) $ is in the image of integer cohomology.  It is
natural to assume that the restriction of~$F$ to~$X^0$, which is closed, also
has a cohomology class which is in the image of integer cohomology.  Then
there exists a map
  $$ A^0\:X^0\longrightarrow \rtpz \tag{3.20} $$
whose differential is~$F$ restricted to~$X^0$ and whose winding number
about~$W_B$ is given by~$q_B/2\pi $.  (These conditions are related by
Stokes' theorem.) 
 
Now we go further and define a gauge potential~$A$ on all of~$X$.  Namely, we
define a gauge potential to be a section 
  $$ A\:X\longrightarrow \scrJ_B \tag{3.21} $$
of~$\scrJ_B$.  In other words, {\it the gauge field is a trivialization of
the magnetic current\/}.  This is a geometric version of~\thetag{3.5}.  In
this picture the gauge field~\thetag{3.21} is not a map to a fixed circle, as
in~\thetag{3.20}, but rather a map to a variable circle defined by the
magnetic current.  The existence of~$A$ forces $\scrJ_B$~to be globally
trivial, just as~\thetag{3.20} forces $[j_B]$~to vanish in absolute
cohomology.  Furthermore, if we trivialize~$\scrJ_B$ on~$X^0$, or even
on~$X\setminus W_B$, then the ratio of~$A$ to the trivialization is the gauge
field~\thetag{3.20}. 
 
We emphasize that it is important that the geometric magnetic
current~$\scrJ_B$ be a particular bundle with connection, not a class of
bundles up to isomorphism.  Otherwise, we could not talk about
trivializations.  Further, once the gauge field is defined as a
trivialization of the magnetic current, then its geometric nature---in
particular its integrality---is tied to that of the magnetic current.  We
remark that this story goes through for $n=\dim X$~arbitrary, except that
$\dim W_B=n-2$ in general.  
 
Suppose now we have electric charges described as the
inclusion~$i_E\:W_E\hookrightarrow X$ of a finite set of oriented points.
The quantization law~\thetag{3.18} asserts that the electric charge is
$\hbar$~times a function 
  $$ q_E\:W_E \longrightarrow \ZZ. \tag{3.22} $$
The exponentiated electric coupling~\thetag{3.17} takes the form
  $$ \prod\limits_{w\in W_E} \exp(iA)^{q_E(w)}. \tag{3.23} $$
If $W_E\cap W_B=\emptyset $ we can use the trivialization to replace~$A$ by
the $\rtpz$-valued function~$A^0$, defined on the complement of~$W_B$, and so
define~\thetag{3.23} as an element of~$\TT\subset \CC$.  Without using the
trivialization, we regard~$\exp(iA)$ as an element of the fiber of the
hermitian line bundle $L_B\to X$ associated to~$\scrJ_B\to X$ via the
character~$x\mapsto \exp(ix)$ of~$\rtpz$.  Then \thetag{3.22}~is an element
of 
  $$ \bigotimes_{w\in W_E}(L_B)^{\otimes q_E(W)}. \tag{3.24} $$
We can imagine a theory in which the electric and magnetic sources $W_E$~and
$W_B$~vary, perhaps as functions of other fields.  Then
\thetag{3.24}~determines a hermitian line bundle with connection over any
parameter space~$T$ of fields.  As explained in~\S{2.1}, this line bundle is
the anomaly in the exponentiated electric coupling.  Note from~\thetag{3.24}
that it is in some sense the integral of the product of the magnetic and
electric currents, but where the currents are refined to
differential-geometric objects~$\jhat_B$ and~$\jhat_E$.  We write the anomaly
in a family $X\to T$ as
  $$ L_{\text{electric}} = \exp\left( 2\pi i\int_{X/T}\jhat_B\cdot \jhat_E
     \right).  \tag{3.25} $$
In this form it generalizes to more complicated situations.

 \subhead \S{3.5} Generalized Differential Cocycles
 \endsubhead

We return briefly to the Hamiltonian formalism, so to a spacetime $M^n =
\EE^1\times N^{n-1}$.  Consider now a general abelian field strength~$F\in
\Omega ^{\dd}(M)$ of multi-degree $\dd=(d_1,\dots ,d_k)$.  Then the currents
$j_B\in \Omega ^{\dd+1}(M)$ and~$j_E\in \Omega ^{n-\dd+1}(M)$ have
corresponding charges~$\bQ_B\in H^{\dd+1}_c(N;\RR)$ and~$\bQ_E\in
H^{n-\dd+1}_c(N;\RR)$.  We generalize Dirac's quantization law~\thetag{3.18}
to postulate full lattices~$\Gbar^\bullet$ in real cohomology so that the
charges~$\Qbar_B,\Qbar_E$ take values in~$\Gbar^\bullet_c(N)$.  The argument
for this follows the one presented in~\S{3.3} for 2-form field strengths.
Namely, the gauge field can only be defined in $\Qbar_B$~is ``quantized'',
and then the electric coupling forces $\Qbar_E$~to be quantized as well.  In
fact, the magnetic current and gauge field should be quantized by one
lattice, and the electric current by a lattice which is complementary in the
sense that there is a duality pairing which generalizes~\thetag{3.18}.
However, to keep things simple we use the same lattice and postulate a
multiplication which implements the analog of~\thetag{3.18}. 
 
The most natural lattice which comes to mind is 
  $$ \Gbar^\bullet(N) = \image\bigl(H^\bullet(N;\ZZ)\longrightarrow
     H^\bullet(N;\RR) \bigr),  $$
but this is not the only possible choice.  Indeed, any generalized cohomology
theory defines a full lattice in real cohomology.   

        \remark{\protag{3.26} {Remark}}
 As an example, consider complex $K$-theory.  Recall that the Chern character
is a ring homomorphism 
  $$ \ch\:K^0(X)\longrightarrow H^{\text{even}}(X;\RR).  $$
The kernel is the torsion subgroup, and the image is a full lattice
isomorphic to~$K^0(X)/\text{torsion}$.  A more canonical version maps to real
cohomology with coefficients in $K^\bullet(pt)\otimes \RR\cong
\RR[[u,u\inv]]$, where $\deg u=2$: 
  $$ \ch\:K^\bullet(X)\longrightarrow H\bigl(X;\RR[[u,u\inv ]]
     \bigr)^\bullet. \tag{3.27} $$
This is a homomorphism of $\ZZ$-{\it graded\/} rings. 
        \endremark

It is natural, now, to refine the charges~$\Qbar_B,\Qbar_E\in
\Gbar^\bullet_c(N)$ to charges~$Q_B,Q_E\in \Gamma ^\bullet_c(N)$ in the
abelian groups, rather than in the groups mod torsion.  This allows for
torsion charges, and in any case $\Gamma (\cdot )$~is better behaved
than~$\Gamma (\cdot )/\text{torsion}$. 

        \remark{\protag{3.28} {Remark}}
 In all examples of which we are aware the lattice~$\Gbar^\bullet$ in real
cohomology is defined by a generalized cohomology theory, which furthermore
is multiplicative.  Locality in quantum field theory suggests that $\Gamma
(X)$~should depend locally on~$X$, and the Mayer-Vietoris property of
generalized cohomology theories is certainly an expression of locality.
However, it seems possible that $\Gamma ^\bullet$~may be defined as something
other than a generalized cohomology theory. 
        \endremark

Switch to the Euclidean setting and drop the support condition.  The
refinement~$\jhat$ of electric and magnetic currents that we seek must encode
both the integral charge~$Q\in \Gamma ^\bullet(X)$ and the local information
of the differential form~$j\in \clf^\bullet(X)$.  A first guess is the fiber
product\footnote{The coefficients should be~$\Gamma ^\bullet(pt)\otimes \RR$,
not simply~$\RR$, but for simplicity we do not incorporate that into the
notation here.} 
  $$ \CD
      A^\bullet_\Gamma (X) @>>> \clf^\bullet(X)\\
      @VVV @VVV \\
      \Gamma ^\bullet(X) @>>> H^\bullet(X;\RR) \endCD $$
Note that $\Qbar$~is the common image of~$Q$ and~$j$ in~$H^\bullet(X;\RR)$.
But the naive fiber product of abelian groups is not sufficient here.
Rather, we need to take the fiber product in the sense of cohomology
theories.  This gives a pullback diagram
  $$ \CD
      \check\Gamma ^\bullet (X) @>>> \clf^\bullet(X)\\
      @VVV @VVV \\
      \Gamma ^\bullet(X) @>>> H^\bullet(X;\RR) \endCD \tag{3.29} $$
in which the upper left-hand entry is the {\it generalized differential
cohomology\/}.  A generalized differential $q$-cocycle is a
triple~$(c,h,\omega )$, where $c$~is a cocycle for~$\Gamma ^q$, $\omega $~is
a closed $q$-form, and $h$~is a $(q-1)$-cocycle for real singular cohomology
whose differential is~$\omega -c$.  In other words, $h$~is a homotopy
from~$c$ to~$\omega $.  There are many possible cocycle models possible for
topological (generalized) cohomology, and they lead to different cocycle
models for the differential theory.  For a heuristic exposition
see~\cite{F5,\S1}; for a detailed development see~\cite{HS}. 
 
Briefly, then, charge quantization is implemented by a choice of generalized
cohomology theory.  The electric and magnetic currents~$j_E,j_B$ are lifted
to generalized differential cocycles~$\jhat_E,\jhat_B$.  The gauge
field~$\check A$ is a trivialization of~$\jhat_B$.  The electric coupling is
written using the product in differential cohomology, and it has an anomaly
computed by~\thetag{3.25}.  This picture of abelian gauge fields is explained
in detail in~\cite{F5,\S2}.

 \subhead \S{3.6} Self-Dual Fields
 \endsubhead

Let $\MM^n$ denote $n$-dimensional Minkowski spacetime, an affine space whose
underlying vector space~$V$ carries a metric of signature~$(1,n-1)$.  The
{\it Poincar\'e group\/} is the double cover of the connected component of
affine isometries of~$\MM^n$.  The corresponding double cover group of the
connected component of linear isometries of~$V$ is the {\it Lorentz group\/}.
In relativistic quantum mechanics a {\it particle\/} is an irreducible
unitary representation of the Poincar\'e group.  We restrict to massless
representations, which may be described as certain function spaces on the
positive nullcone~$N^+$ in the dual space~$V^*$.  The Lorentz group acts
on~$N^+$, transitively if~$n\ge3$, and the reductive part of the
stabilizer---the {\it massless little group\/}---is isomorphic to the compact
group~$\Spin(n-2)$.  Massless particles correspond to irreducible complex
representations of~$\Spin(n-2)$.
 
Our interest is in the exterior power representations~${\tsize\bigwedge} ^p$.
They factor through~$SO(n-2)$, so correspond to {\it bosons\/}.  The Hodge
$*$~operator gives an isomorphism
  $$ {\tsize\bigwedge} ^p\cong {\tsize\bigwedge} ^{n-2-p}, \tag{3.30} $$
so only the representations for $p=1,\dots ,\left[ \frac{n-2}{2} \right] $
are distinct.  If $n=4\ell +2,\;n\ge6$, the representation~${\tsize\bigwedge}
^{2\ell }$ splits into the sum of two irreducible representations
  $$ {\tsize\bigwedge} ^{2\ell } \cong {\tsize\bigwedge} ^{2\ell }_+ \oplus
     {\tsize\bigwedge} ^{2\ell }_- \tag{3.31} $$
In quantum field theory there is a restriction on the total particle
representation---the CPT~theorem---which for $n$~even asserts that the
representation of the little group must carry a real structure, which is why
we only consider the decomposition~\thetag{3.31} for~$n\equiv 2\pmod4$.  The
irreducible representation~${\tsize\bigwedge} ^{2\ell }_+$ gives the {\it
self-dual\/} particle; ${\tsize\bigwedge} ^{2\ell }_-$~the anti-self-dual
particle.  For~$n=2$ the representation~${\tsize\bigwedge} ^0$ is of course
irreducible, but the positive nullcone~$N^+$ breaks into two orbits
$N^+=N^+_L\cup N^+_R$ of the Lorentz group, and the massless representation
of~Poincar\'e corresponding to~${\tsize\bigwedge} ^0$ breaks up as a sum of
two irreducible representations, one supported on~$N^+_L$ and one on~$N^+_R$.
 
Free field theories on~$\MM^n$ lead to particles by quantization: the space
of solutions to the Poincar\'e-invariant free field equations is a real
symplectic vector space with a symplectic action of Poincar\'e.  Its
quantization is the Hilbert space completion of the symmetric algebra of a
unitary representation of~Poincar\'e.  This representation is the {\it
particle content\/} of the theory.  For example, the classical Maxwell
equations~\thetag{3.2} with~$j_E=0$ lead to the massless irreducible
representation~${\tsize\bigwedge} ^1$.  (See~\cite{F6,Lecture~3} for one
account.)  More generally, we may consider a {\it real\/} $d$-form field
strength~$F$ on~$\MM^n$ which satisfies the free field equations 
  $$ dF=d*F=0. \tag{3.32} $$
This free system has a lagrangian description in terms of a gauge
potential~$A$ up to gauge transformations; the relevant quotient space of
fields is~\thetag{3.10} with 1-forms replaced by $(d-1)$-forms.  The
particle content of the corresponding quantum system is the massless
representation~${\tsize\bigwedge} ^{d-1}$.  The symmetry~\thetag{3.30} of
particles corresponds to the {\it electromagnetic duality\/} symmetry 
  $$ F \longleftrightarrow *F  $$
of the classical field equations~\thetag{3.32}---see \theprotag{3.4}
{Remark}. 
 
Specialize to~$n=4\ell +2$ and~$d=2\ell +1$.  Then the Hodge~$*$ fixes the
space~$\Omega ^{2\ell +1}(\MM)$ of $d$-forms on~$\MM^n$, and since~$*^2=1$ we
have a decomposition
  $$ \Omega ^{2\ell +1}(\MM^n) \cong \Omega _+^{2\ell +1}(\MM^n) \oplus
     \Omega _-^{2\ell +1}(\MM^n)  $$
This induces a decomposition of the space of solutions to~\thetag{3.32} into
a direct sum of two symplectic subspaces, each invariant under Poincar\'e, so
each giving a free relativistic classical mechanical system in its own right.
In fact, the quantization of these systems give the irreducible massless
representations~${\tsize\bigwedge} ^{2\ell }_\pm$ of Poincar\'e.  (For~$n=2$
we obtain the representations supported on~$N^+_L,N^+_R$.)  In other words,
the classical systems which correspond to (anti-)self-dual particles are the
$(2\ell +1)$-forms which satisfy ~\thetag{3.32} and
  $$ F = \pm *F. \tag{3.33} $$

In~\S{3.2} we explained how the field equations~\thetag{3.32} arise from an
action principle on gauge potentials.  Unfortunately, there is no
straightforward Poincar\'e-invariant lagrangian theory which also gives the
(anti-)self-duality equation~\thetag{3.33}.  This is not a handicap for the
classical theory---the field equations contain all the information---but
leads to an extra complication in the {\it Euclidean\/} version of the
quantum theory.  Namely, to define Euclidean correlation functions on
arbitrary Riemannian manifolds, there is an extra piece of data which is
needed~\cite{W5}.  It is best expressed as a quadratic refinement of the
bilinear form~\thetag{3.25} on electric and magnetic currents.  Dirac
quantization enters as for any gauge field, so the quadratic form is best
described in the language of generalized differential cocycles.  Note that
for (anti-)self-dual fields the magnetic current determines the electric
current, and the Euclidean partition function on~$X$ depends on this single
current.  A precise geometric formula for this partition function and its
anomaly is still lacking in general.

 \subhead \S{3.7} Ramond-Ramond Charge and $K$-Theory
 \endsubhead

The physically correct choice of a generalized cohomology theory~$\Gamma $
for the quantization of charge is motivated by many different considerations
in the quantum theory.  We illustrate this in Type~II superstring theory,
where integer cohomology quantizes Neveu-Schwarz~(NS) charge and complex
$K$-theory quantizes Ramond-Ramond~(RR) charge.  Variations of these
arguments occur in other superstring theories.
 
We work with the field theory approximation to the low energy limit of
Type~II superstring theory.  The first step is the quantization of the free
superstring~\cite{d'H,\S7.8} on~$\MM^n$.  The massless particles which occur
include exterior power representations~${\tsize\bigwedge} ^p$ of the little
group~$\Spin(8)$.  As explained in~\S{3.6} they are obtained by quantizing
differential form fields.  Specifically, there is a 3-form NS~field strength
and there are RR~field strengths of various degrees.  In Type~IIA these have
degrees~2 and~4; in Type~IIB the degrees are~1, 3, and~$5_+$, where the
subscript indicates that the field is self-dual.  (The self-dual field
corresponds to the representation~${\tsize\bigwedge} ^4_+$ of~$\Spin(8)$.)
We need to specify the appropriate Dirac quantization condition for these
gauge fields.  We work in the Euclidean theory, so on a Riemannian
10-manifold~$X$.
 
We begin with the NS 2-form gauge potential, which we denote~$B$.  It turns
out that the correct quantization condition is integer cohomology.  In other
words, in the absence of magnetic current $B$~is lifted to a cocycle~$\Bh$
for the ordinary differential cohomology group~$\Hh^3(X)$.  There are a few
justifications for using integer cohomology here.  First, the $B$-field has
an electric coupling of the form~\thetag{3.17}.  The fundamental string is
represented by a map~$\phi \:\Sigma \to X$ of a closed oriented
surface~$\Sigma $ to~$X$, and the exponentiated electric coupling is
  $$ \exp\left( \frac{i}{2\pi \alpha '\hbar}\int_{\Sigma }\phi ^*B \right),
     \tag{3.34} $$
where $\alpha '$~is the {\it Regge slope\/}.  Comparison with~\thetag{3.17}
shows that the electric NS charge of the fundamental string is~$q_E=1/(2\pi
\alpha ')$.  Now $\Sigma $~carries a fundamental class in integer cohomology,
and this allows us to lift~\thetag{3.34} to an integral of an ordinary
differential cocycle~$\Bh$:
  $$ \exp\left( \frac{i}{2\pi \alpha '\hbar}\int_{\Sigma }\phi ^*\Bh \right).
      \tag{3.35} $$
This simple form of the electric coupling is the first rationale for using
integer cohomology to quantize the NS~charge; it is not clear how to
write~\thetag{3.35} with other differential cohomology theories.  This
electric coupling occurs in the worldsheet theory, where $\Bh$~is a
background field and $\phi $~a dynamical field.  In the strong coupling limit
of Type~II superstring theory there can be a fixed background source~$\phi
:W\hookrightarrow X$ for the dynamical field~$\Bh$, which is more in keeping
with the context of~\thetag{3.17}.  In that situation the dual magnetically
charged object is the NS~5-brane, whose magnetic charge is an integer
multiple of~$(2\pi )^2\alpha '\hbar$, in accordance with~\thetag{3.18}.  This
quantization of charge is consistent with the form of classical 5-brane
solutions to supergravity~\cite{CHS,\S6}, \cite{P1,\S\S14.1,14.4}, and
provides further confirmation that integer cohomology is the correct choice
of quantization law for the $B$-field.\footnote{Contrast to the D-branes
which couple to the RR~gauge fields.  In that case the charge is represented
by a complex vector bundle on the D-brane, not by an integer, and this bundle
appears in the electric coupling.  The low energy fluctuations of the
NS~5-brane include a gauge field living on the brane, but it is not involved
in the electric coupling.}  Another piece of evidence that integer cohomology
is correct involves twistings of RR-fields, as explained below.
 
We turn now to the RR~gauge fields, where the correct quantization is
provided by complex $K$-theory if the NS~$\Bh$-field vanishes.  Before
reviewing physical arguments for this, we interpret the statement
mathematically in terms of differential $K$-theory.
(Compare~\cite{MW,(2.17)}.)  The first step is to introduce electromagnetic
duals to the minimal set of differential form fields.  So we postulate
inhomogeneous RR~field strengths
  $$ G' = \cases G_2 + G_4 + G_6 + G_8, &\text{Type~IIA};\\G_1 + G_3 + G_5 +
     G_7 + G_9,&\text{Type~IIB},\endcases  $$
where $G_d$~is a differential form of degree~$d$.  In the classical theory on
~$\MM^{10}$ we impose the self-duality equation~$G'=*G'$, but in the
Euclidean quantum theory on a Riemannian 10-manifold the self-duality
condition is manifest by using a certain quadratic form in the definition of
the partition function (see~\S{3.6}).  It is convenient to redefine the
RR~field strength to be homogeneous.  Recalling from~\theprotag{3.26}
{Remark} the inverse Bott element~$u\in K^2(pt)$, we set
  $$ G = \cases u\inv G_2 + u^{-2}G_4 + u^{-3}G_6 + u^{-4}G_8,
     &\text{Type~IIA};\\u^{-1}G_1 + u^{-2}G_3 + u^{-3}G_5 + u^{-4}G_7 +
     u^{-5}G_9,&\text{Type~IIB},\endcases  $$
Then $\deg G=0$ in~IIA and $\deg G=-1$ in~IIB.  In the absence of current,
then, the refined gauge field~$\Ch$ is a cocycle for an element
of~$\Kh^\bullet(X)$, where $\bullet=0$ in~IIA and $\bullet=-1$ in~IIB.  Note
that the defining pullback square~\thetag{3.29} for differential $K$-theory
is
  $$ \CD
      \Kh ^\bullet (X) @>>> \clf\bigl(X;\RR[[u,u\inv ]]\bigr)^\bullet\\
      @VVV @VVV \\
      K^\bullet(X) @>>> H\bigl(X;\RR[[u,u\inv ]]\bigr)^\bullet \endCD
      $$
The bottom map is the Chern character~\thetag{3.27}.  The current~$\jhat$ is
a cocycle of one degree higher than~$\Ch$; if it is nonzero, then $\Ch$~is a
trivialization of~$\jhat$. 
 
One indication that $K$-theory is the correct cohomology theory to quantize
RR~charge is the form of RR-charged objects.  These are {\it D-branes\/};
see~\cite{P2}, \cite{P1,\S13} for a review.  Geometrically, a D-brane in the
Euclidean theory is a submanifold $i\:W\hookrightarrow X$ of codimension~$r$
together with a complex vector bundle $Q\to W$ and unitary connection.  In
Type~IIA the codimension~$r$ is odd, in Type~IIB it is even.  The vector
bundle~$Q$ is analogous to the function~\thetag{3.22} in electromagnetism; it
encodes the charge carried by the D-brane.  Physicists refer to~ $\rank Q$ as
the number of D-branes, so a single D-brane comes with a line bundle and
unitary connection, that is, a 1-form gauge field quantized by integer
cohomology.  We allow virtual bundles as well; negative rank corresponds to
D-antibranes.  One motivation for the vector bundle~$Q\to W$ is the
Chan-Paton construction in {\it open\/} string theory~\cite{d'H,\S2.11}.  The
important point for us is the coupling of the D-brane to the RR~gauge
potential~$C$ involves the gauge field on the brane.  This was discussed in
several papers, for example~\cite{GHM}, \cite{MM}, \cite{CY}.  In particular,
motivated by the presence of the $\Ahat$-genus in the differential form
expression of the coupling, Minasian and Moore~\cite{MM} suggested the
$K$-theory formula for RR~charge.  It is important to note that this coupling
has the standard form~\thetag{3.17} when the RR~gauge potential is lifted to
a differential $K$-theory cocycle~$\Ch$: 
  $$ \exp\left( \frac{i}{2\hbar}\int_{W} u^{-\left[ \frac r2 \right] }\;
     \Ch\cdot \overline{\Qh} \right).   $$
Here the bundle~$Q\to W$ with its connection gives rise to a differential
$K$-theory cocycle~$\Qh$ whose complex conjugate is~$\overline{\Qh}$.  The
factor of~$1/2$ is inserted since $\Ch$~is self-dual, and the correct
interpretation involves the quadratic form mentioned above.\footnote{The
notion of a lagrangian for self-dual fields is itself only a schematic for
the definition of the partition function; see the end of~\S{3.6}.}  The
RR~charge of the D-brane is~$u^{-\left[ \frac r2 \right] }
i_*[\overline{Q}]$, where $i_*\:K^0(W)\to K^r(X)$ is the pushforward, which
is defined if the normal bundle to~$W$ is oriented and $\text{spin}^c$.  In
fact, there is a fermion anomaly in the perturbative open string~\cite{FW}
which shows that for $\rank Q=1$ the line bundle~$Q$ should be interpreted in
terms of $\text{spin}^c$ structures.  (In general one must interpret it in
{\it twisted\/} $K$-theory.)  This anomaly, which fits the orientation
condition for $K$-theory, is yet another piece of evidence for the
quantization condition.
 
A rather different piece of evidence was put forth by Witten~\cite{W4}.  He
exhibited objects in superstring theory with torsion RR~charge which
naturally live in $K$-theory (rather than integer cohomology, for example).
His examples lie in Type~I, but Type~I may be constructed from Type~II and so
his examples are also relevant for understanding Type~II.  Furthermore,
Witten gave several formal arguments for $K$-theory and also showed that
Sen's brane-antibrane annihilation~\cite{Se} may be interpreted as the
difference construction in $K$-theory.
 
The choices of integer cohomology to quantize the NS~$\Bh$-field and
$K$-theory to quantize the RR~$\Ch$-field fit together if both are nonzero.
Namely, the $\Bh$-field provides a {\it twisting\/} of differential
$K$-theory, and the $\Ch$-field lives in this twisted theory.  First,
topologically a cocycle for~$H^3(X)$ defines a twisting of $K$-theory, and if
$[\Bh]\in H^3(X)$ is nonzero then the RR~charge lives in this twisted
$K$-theory.  There is a more refined statement on the level of differential
cohomology: $\Ch$~is a $\Bh$-twisted differential $K$-theory cocycle.  We
remark that the mathematical groundwork for these twisted theories has not
yet been fully developed.  Also, these assertions about twisted
(differential) $K$-theory enter the anomaly considerations to which we now
turn.

The Type~II theories on a 10-manifold are believed to be anomaly-free.  For
Type~IIA this is easily proved, since both chiralities of fermion occur and
so the fermion anomalies cancel by an elementary argument; see the footnote
preceding~\thetag{A.29}.  In Type~IIB the {\it local\/} fermion anomalies
cancel the local anomaly in the self-dual RR~gauge field, that is, the total
partition function is a section of a line bundle with zero curvature.
However, as mentioned at the end of~\S{3.6}, there is not yet a precise
definition of the partition function of the self-dual field which
demonstrates a cancellation of {\it global\/} anomalies, i.e., which
demonstrates that the line bundle has no holonomy (and even better has a
canonical flat section of unit norm).  Nevertheless, that $K$-theory
quantizes the RR~gauge field must enter into the eventual argument.  What has
been worked out are the additional anomalies due to a D-brane~\cite{FH},
~\cite{MW}.  Here the electric coupling anomaly~\thetag{3.25} cancels a
fermion anomaly~\thetag{2.8}.  The formulas for both live, at least
conjecturally, in differential $K$-theory, and if the RR~gauge field were
lifted to a different generalized differential cohomology theory this
argument would not work. 
 
Similar reasoning indicates that in any situation where this {\it
Green-Schwarz mechanism\/} cancels fermi anomalies against and electric
coupling anomaly, the gauge field will be quantized by a cohomology theory
which, if not some form of $K$-theory, at least maps to it.

 \newpage
 \head
 Appendix: The Partition Function of Rarita-Schwinger Fields\\ \\ 
Daniel S. Freed\\Jerome A. Jenquin
 \endhead
 \comment
 lasteqno A@ 29
 \endcomment

This appendix grew out of a concrete question: Why for the computation of
anomalies is the Rarita-Schwinger field on an even dimensional manifold~$X$
treated as a spinor field coupled to the virtual bundle~$TX - 1$, whereas in
11~dimensional M-theory it is treated as a spinor field coupled to~$TX - 3$?
The perturbative anomaly computation in even dimensions is explained
in~\cite{AgW}, and the statement in 11~dimensions appears
in~\cite{DMW,(2.7)}.  Here we analyze general Rarita-Schwinger fields from
first principles.  First, we quantize the free classical Rarita-Schwinger
field on Minkowski spacetime to obtain particle representations of the
Poincar\'e group.  There are three different classical theories we consider,
and they lead to different particle content.  The first of these generalizes
neatly to Euclidean field theory, and using it we define the Rarita-Schwinger
partition function.  This leads to the anomaly computations cited above.

The material here is well-known to physicists.  Our notation follows the
treatments in~\cite{DF,\S3} and~\cite{F6,Lecture~3} of scalar, spinor, and
1-form fields.

 \subhead \S{A.1}. Quantization of Spinor Fields: Review
 \endsubhead

Let $\MM^n$ denote $n$-dimensional Minkowski spacetime.  It is an affine space
modeled on a vector space~$V$ equipped with a Lorentz metric~$g$ of
signature~$(1,n-1)$.  Fix a basis~$\{e_\mu \}$ of~$V$ and dual basis~$\{e^\mu
\}$ of~$V^*$.  These give rise to coordinates on~$\MM^n$, so partial
derivatives~$\partial _\mu $.  Let $\voll$~denote the density induced from
the metric.  When we take Fourier transforms below we choose a basepoint
of~$\MM^n$ to identify it with~$V$; then the Fourier transform is a function
on~$V^*$.  Let~$N\subset V^*$ be the cone of null covectors, and~$N^+\subset
N$ a distinguished component of~$N\setminus \{0\}$.  (This gives a notion of
positive energy.)

Let~$\Spin(V)$ denote the connected Lorentz spin group.  We develop the
theory for an arbitrary finite-dimensional real spinor representation~$S$
of~$\Spin(V)$.  It is a fact~\cite{De,Chapter~6} that for any~$S$ there exist
symmetric pairings
  $$ \alignedat2
      \Gamma &\: S^*\otimes S^*&&\longrightarrow V \\
      \tilde \Gamma &\: S\;\otimes \;S&&\longrightarrow V\endaligned
      $$
which satisfy a Clifford relation.  Fix bases~$\{f^a\}$ of~$S$ and dual
basis~$\{f_a\}$ of~$S^*$.  Then the Clifford relation reads 
  $$ \Gamma ^\mu _{a b }\tilde{\Gamma }^{\nu b c } + \Gamma ^\nu _{a b
     }\tilde{\Gamma }^{\mu b c } = 2g^{\mu \nu }\delta ^c _a ,  $$
where $g^{\mu \nu }$~is the inverse metric~$g\inv $ on~$V^*$.  Put
differently, $\Gamma, \tilde\Gamma $~determine Clifford multiplications
  $$ \aligned
      c&\: V^* \longrightarrow \Hom(S,S^*) \\
      c&\: V^* \longrightarrow \Hom(S^*,S)\endaligned \tag{A.1} $$
which satisfy 
  $$ c(k)c(\ell ) + c(\ell )c(k) = 2g\inv ( k,\ell ),\qquad k,\ell \in
     V^*. \tag{A.2} $$
Thus $S\oplus S^*$~is a real Clifford module for the Clifford
algebra~$\Cliff(V^*,g\inv )$. 
      
Spinor fields and Rarita-Schwinger fields are odd in the sense of
$\zt$-graded geometry.  Let $\Pi S$~denote the odd vector space which is the
parity-reversal of~$S$.\footnote{More precisely, $\Pi
S=\CC^{\text{odd}}\otimes S$, where $\CC^{\text{odd}}$~is a fixed odd line.}
Then a spinor field is a function 
  $$ \psi \:\MM^n\longrightarrow \Pi S. \tag{A.3} $$
The lagrangian is the symmetric (in the graded sense) bilinear form based on
the Dirac operator: 
  $$ L = \frac 12 \psi \Dirac\psi \;\voll = \frac 12 \tilde\Gamma^{\mu ab}\psi
     _a\partial _\mu \psi _b\;\voll.  \tag{A.4} $$
Now the Fourier transform of~\thetag{A.3}
  $$ \hpsi\:V^*\longrightarrow \Pi \SC  $$
lands in the complexified spin space and satisfies the reality condition
  $$ \hpsi(-k) = \overline{\hpsi(k)}. \tag{A.5} $$
The equation of motion derived from~\thetag{A.4} is the Dirac equation 
  $$ c(k)\hpsi(k) = \tilde\Gamma^{\mu ab}k_\mu \psi_a(k)f_b = 0. \tag{A.6} $$
If $|k|^2\not= 0$ we apply~$c(k)$ to~\thetag{A.6} to deduce that $\hpsi$~is
supported on the nullcone~$N$.  Define a complex vector bundle $S'\to
N\setminus \{0\}$ by
  $$ S'_k=\ker\bigl(c(k)\:\SC\to\SC^* \bigr),\qquad |k|^2=0,\quad k\not= 0
     \tag{A.7} $$
Typically this is a bundle of rank~$\frac 12\dim S$, though there are
exceptions in $n=2$~dimensions.  Using~\thetag{A.6} we
identify\footnote{Throwing out $k=0$ does not cause problems if~$n\ge3$, but
leads to special considerations in dimension~$n=2$.} the space of classical
solutions with the space of sections of~$\Pi S'\to N\setminus \{0\}$ which
satisfy the reality condition~\thetag{A.5}.  The lagrangian determines a
symplectic structure on this vector space.
 
Quantization proceeds by complexifying the symplectic space of solutions---so
dropping the reality condition---and then considering the lagrangian subspace
of sections supported on~$N^+\subset N$.  This is a unitary representation of
the Poincar\'e group called the {\it one-particle Hilbert space\/}; the full
quantum Hilbert space is the completion of its symmetric algebra (in the
graded sense).
 
We describe the bundle~$S'\to N^+$ more explicitly as follows.  Introduce the
real vector bundle $V'\to N^+$ of rank~$n-2$ whose fiber at~$k\in N^+$ is
  $$ V'_k = \frac{k^{\perp}}{\RR\cdot k}.  $$
The Lorentz group~$\Spin(V)$ acts on~$N^+$ and there is a lifted action on
the bundles~$V',S'$.  The subgroup of~$\Spin(V)$ which stabilizes~$k$ has a
reductive part which is~$\Spin(V'_k)$, also known as the ``little group''.
It is isomorphic to the compact spin group~$\Spin(n-2)$.  These groups fit
together to form a bundle of groups $\Spin(V')\to N^+$.  Then $S'_k$~is a
complex spin representation of~$\Spin(V'_k)$, and the associated unitary
representation of Poincar\'e is called a massless spin~1/2 particle.
(See~\cite{De,Chapter~5} for more about the representation~$S'$.)  Depending
on~$S$ this may or may not be irreducible, so is better called a collection
of spin~1/2 particles.  The complexification of $V'\to N^+$ is also
associated to a unitary representation of Poincar\'e, irreducible if~$n\ge3$,
called the spin~1 particle.

 \subhead \S{A.2}. Quantization of Rarita-Schwinger Fields: First Approach
 \endsubhead

Continuing the discussion of particles in relativistic quantum mechanics, we
first describe a spin~3/2 particle.  As in~\thetag{A.7} define a complex
vector bundle $S''\to N^+$ by
  $$ S''_k=\ker\bigl(c(k)\:\SC^*\to\SC \bigr),\qquad k\in N.  $$
Then Clifford multiplication~\thetag{A.1} induces a map 
  $$ c'\:V'\otimes S'\longrightarrow S'' \tag{A.8} $$
which is equivariant for~$\Spin(V')$.  Set $R'=\ker c'$.  The sections of the
odd $\Spin(V)$-equivariant complex vector bundle $\Pi R'\to N^+$ form a
unitary representation of the Poincar\'e group which represents a collection
of spin~3/2 particles.  (Again, depending on~$S$ it may or may not be
irreducible.)
 
It is useful to note that there is a splitting of~\thetag{A.8}: 
  $$ \aligned
      S'' &\longrightarrow V'\otimes S' \\
      s^af_a &\longmapsto \frac 1n g_{\mu \nu }\Gamma ^\mu _{ab}s^ae^\nu
     \otimes f^b.\endaligned  $$
Thus we have a decomposition 
  $$ V'\otimes S' \cong R'\oplus S'' \tag{A.9} $$
as $\Spin(V)$-equivariant complex vector bundles over~$N^+$.
 
A {\it Rarita-Schwinger field\/} is a fermionic field 
  $$ \chi \:\MM^n\longrightarrow V^*\otimes \Pi S.  $$
We write $\chi =\chi _\mu e^\mu =\chi _{\mu a}e^\mu \otimes f^a$, where $\chi
_\mu \in \Pi S$ and $\chi _{\mu a}\in \Pi \RR$.  The first lagrangian we
consider generalizes~\thetag{A.4}: 
  $$ L = \frac 12\chi \Dirac\chi =\frac 12 \tilde\Gamma ^{\mu ab}g^{\nu \rho
     } \chi _{\nu a}\partial _\mu \chi _{\rho b}. \tag{A.10} $$
Note the factor of the inverse metric to absorb the vector index on~$\chi $.
The associated Euler-Lagrange equation, written on the Fourier transform, is
  $$ \bigl(id\otimes c(k) \bigr)\hchi(k)= \tilde\Gamma ^{\mu ab}k_\mu
     \hchi_{\rho a}(k)f_b = 0. \tag{A.11} $$
As before we conclude that $\hchi$~is supported on the nullcone~$N$.
Polarize---drop the reality condition and restrict to~$N^+\subset N$---to
obtain the space of sections of $\VC^*\otimes \Pi S'\to N^+$.  Now the
$\Spin(V)$-equivariant vector bundle $V^*\to N^+$ decomposes as the sum
of~$V'$ and a rank~2 trivial bundle.  Hence, using~\thetag{A.9} we see that
the one-particle Hilbert space of the theory is the space of sections of
  $$ \Pi \bigl(R'\oplus S''\oplus S'\oplus S' \bigr) \longrightarrow
     N^+. \tag{A.12} $$
In other words, the particle content of the theory is a collection of
spin~3/2 particles based on~$R'$, a collection of spin~1/2 particles based
on~$S''$, and two collections of spin~1/2 particles based on~$S'$.

        \remark{\protag{A.13} {Remark}}
 To obtain a collection of pure spin~3/2 particles we need to ``subtract''
off the spin~1/2 particles in~\thetag{A.12}.  There are various mechanisms to
do so.  Our interest is in the Euclidean partition function, and it is in
that context that we isolate the spin~3/2 particles.
        \endremark

        \remark{\protag{A.14} {Remark}}
 There is an analogous quantization of a 1-form field on Minkowski spacetime
which leads to a spin~1 particle and two spin~0 particles.  The more usual
quantization of 1-forms uses a gauge symmetry to avoid the spin~0 particles.
We give the analogous treatment of Rarita-Schwinger fields in the next
section.
        \endremark

 \subhead \S{A.3}. Quantization of Rarita-Schwinger Fields: Second Approach
 \endsubhead

We begin with some linear algebra.  Recall from~\thetag{A.1} that $S\oplus
S^*$ is a real Clifford module for~$\Cliff(V^*)$.  As a real vector space
$\Cliff(V^*)$ is isomorphic to~${\tsize\bigwedge} ^\bullet(V^*)$, so there is
an induced map
  $$ \ct\:{\tsize\bigwedge} ^3V^* \longrightarrow \End(S\oplus S^*)
      $$
which exchanges~$S$ and~$S^*$.  From the Clifford identity~\thetag{A.2} it
follows that for $k_1,k_2,k_3\in V^*$ we have
  $$ \ct(k_1\wedge k_2\wedge k_3) = c(k_1)c(k_2)c(k_3) - \langle k_1,k_2
     \rangle c(k_3) + \langle k_1,k_3 \rangle c(k_2) - \langle k_2,k_3
     \rangle c(k_1), \tag{A.15} $$
where we write~$\langle \cdot ,\cdot \rangle$ for the inverse metric
on~$V^*$.
 
We use~$\ct$ in the following lagrangian for a Rarita-Schwinger field $\chi
\:\MM^n\to V^*\otimes \Pi S$: 
  $$ L = \frac 12 \chi \ct\partial \chi \;\voll = \frac 12 \chi _\mu
     \ct{}^{\mu \nu \rho }\partial _\nu \chi _\rho \;\voll. \tag{A.16} $$
This lagrangian has a gauge invariance.  Namely, if $\psi \:\MM^n\to \Pi S$ is
a spinor field, then $\partial \psi $~is a Rarita-Schwinger field and
$L(\partial \psi )=0$.  So in this model we consider the space of fields to
be the quotient space 
  $$ \frac{\{\chi \:\MM^n\to V^*\otimes \Pi S\}}{\partial \{\psi \:\MM^n\to \Pi
     S\}}.  \tag{A.17} $$
The solutions to the equation of motion on this quotient form a symplectic
vector space, and this is what we quantize to obtain the free quantum theory.
 
Now the  equation of motion  derived from~\thetag{A.16} is  $\ct\partial \chi
=0$.  The maps
  $$ \SC @>A_k>> \VC^*\otimes \SC @>B_k>> \VC\otimes \SC^*  $$
defined by 
  $$ \aligned
      A_k(s) &= k\otimes s \\
      B_k(\ell \otimes s) &= e_\mu \otimes \ct(e^\mu \wedge k\wedge \ell
     )s\endaligned  $$
satisfy $B_k\circ A_k=0$.  The equation of motion on the Fourier transform
asserts $\hchi(k)\in \ker B_k$. 

        \proclaim{\protag{A.18} {Lemma}} 
 {\rm (i)}\ If $|k|^2\not= 0$ then $\ker B_k = \im A_k$.
 {\rm(ii)}\ If $|k|^2=0$ and $k\not= 0$, then
  $$ \frac{\ker B_k}{\im A_k} \cong R'_k.$$
        \endproclaim

\flushpar
 It follows that we can identify the space of solutions on the quotient
space~\thetag{A.17}, after complexifying and polarizing, with the space of
sections of $R' \to N^+$.  In other words, in this approach we obtain exactly
the desired collection of spin~3/2 particles.

        \demo{Proof}
 For~(i) choose the basis $e_1,\dots ,e_n$ of~$V$ to be orthogonal and
have~$e_1=k$.  Assume that $\sum\limits_{i>1}e^i\otimes s^i,\;s^i\in \SC$, is
in~$\ker B_k$; we must show that~$s^i=0$.  The hypothesis implies 
  $$ \sum\limits_{i>1}\ct(e^\mu \wedge e^1\wedge e^i)s^i=0,\qquad \mu
     =1,\dots ,n.  $$
For~$\mu =1$ we learn nothing.  If~$\mu >1$, then from~\thetag{A.15} we have 
  $$ \split
      &= \sum\limits_{i>1}\left[ c(e^\mu )c(e^1)c(e^i) + g^{\mu i}c(e^1)
     \right]s^i \\
      &= \sum\limits_{i>1} -c(e^\mu )c(e^i)c(e^1)s^i \;+\; g^{\mu \mu
     }c(e^1)s^\mu \\
      &= \sum\limits_{{i\not= \mu }\atop{i>1}} c(e^\mu
     )c(e^1)c(e^i)s^i.\endsplit \tag{A.19} $$
Apply~$c(e^1)c(e^\mu )$ and set $t^i=c(e^i)s^i$; then 
  $$ \sum\limits_{{i\not= \mu }\atop{i>1}} t^i=0,\qquad \mu =2,\dots
     ,n. \tag{A.20} $$
Adding these equations we deduce $\sum\limits_{i>1}t^i=0$, and combining
with~\thetag{A.20} we find $t^\mu =c(e^\mu )s^\mu =0$ for all~$\mu $.
Apply~$c(e^\mu )$ to conclude~$s^\mu =0$. 
 
For~(ii) choose the basis of~$V$ to have~$e_1=k$ and $g_{11}=g_{22}=0$,
$g_{12}\not= 0$, $g_{1i}=g_{2i}=g_{ij}=0$, and~$g_{ii}\not= 0$,
where~$i,j>2$, $i\not= j$.  First, we show that $B_k(e^2\otimes s)=0,\;s\in
\SC$, implies~$s=0$.  For then if~$\mu >2$ we apply~\thetag{A.15} to find
  $$ 0=\ct(e^\mu \wedge e^1\wedge e^2)s = \left[ c(e^\mu )c(e^1)c(e^2) -
     g^{12}c(e^\mu ) \right]s, \tag{A.21} $$
from which $c(e^1)c(e^2)s = g^{12}s$, since $c(e^\mu )^2=g^{\mu \mu }\not=
0$.  Apply~$c(e^1)$ to deduce $s\in \ker c(e^1)$.  Apply~$c(e^2)$ to deduce
$s\in \ker c(e^2)$.  Then since $c(e^1)c(e^2) + c(e^2)c(e^1)=2g^{12}\not= 0$
we find~$s=0$. 
 
Finally, suppose $\sum\limits_{i>2}e^i\otimes s^i,\;s^i\in \SC$, is in~$\ker
B_k$.  Then 
  $$ \sum\limits_{i>2}\ct(e^\mu \wedge e^1\wedge e^i)s^i=0,\qquad i=1,\dots
     ,n. \tag{A.22} $$
For~$\mu =1$ we learn nothing.  Set~$\mu =2$ and apply~\thetag{A.15} to find 
  $$ c(e^2)c(e^1)\sum\limits_{i>2}c(e^i)s^i =
     g^{12}\sum\limits_{i>2}c(e^i)s^i.  $$
As in the argument following~\thetag{A.21} we conclude
  $$ \sum\limits_{i>2}c(e^i)s^i = 0. \tag{A.23} $$
For~$\mu >2$ we find from~\thetag{A.22}, analogously to~\thetag{A.19}, that 
  $$ c(e^1)\sum\limits_{{i\not= \mu }\atop{i>2}}c(e^i)s^i = 0.  $$
Together with~\thetag{A.23} this implies $c(e^\mu )s^\mu \in \ker c(e^1)$,
whence
  $$ s^\mu \in \ker c(e^1). \tag{A.24} $$
Then \thetag{A.23} and~\thetag{A.24} together imply
$\sum\limits_{i>2}e^i\otimes s^i\in R'$, as desired.
        \enddemo

 \subhead \S{A.4}. Quantization of Rarita-Schwinger Fields: Third Approach
 \endsubhead
 
Define the real representation~$R$ of~$\Spin(V)$ as the kernel of Clifford
multiplication $V^*\otimes S\to S^*$.  In this approach we take a
Rarita-Schwinger field to be a map
  $$ \tau \:\MM^n \longrightarrow \Pi R.  $$
Then it is easy to verify using~\thetag{A.15} that the
lagrangians~\thetag{A.10} and~\thetag{A.16} agree when $\tau $~replaces~$\chi
$.  There is no gauge symmetry in this approach.  The classical equations may
be analyzed as in the text following~\thetag{A.11}.  Thus the relevant
equivariant vector bundle over~$N^+$ is the parity-reversed kernel of
Clifford multiplication $\VC^*\otimes S'\to \SC^*$.  This is isomorphic to
$\Pi \bigl(R'\oplus S' \bigr)\to N^+$.  So in this approach we obtain a
collection of spin~3/2 particles and a collection of spin~1/2 particles.

 \subhead \S{A.5}. The Euclidean Partition Function of a Rarita-Schwinger Field
 \endsubhead

Here the setting is a Riemannian manifold~$X$.  One defines fields, a
lagrangian, and an action as on Minkowski spacetime, but the classical
equations of motion no longer carry physical meaning.  Rather, these data are
inputs into the functional integral; see~\thetag{2.2}.  The fields and action
are functorial in~$X$.  The correlation functions on Euclidean space are
meant to have analytic continuations to Minkowski spacetime, and in this way
one produces a relativistic quantum mechanical theory.  Our goal, then, is to
produce Euclidean theories---fields and an action---which analytically
continue to the quantum theory of the spin~3/2 particle based on~$R'$.  We
require that $X$~be compact and spin (the compactness may be relaxed if
conditions at the ends are imposed), and focus on the partition function
rather than on general correlation functions.
 
In~\S2 we indicated that the choice of a real spin representation~$S$ of the
Lorentz spin group~$\Spin(V)$---which we use to construct spinor fields
$\MM^n\to \Pi S$ on Minkowski spacetime and then a collection of spin~1/2
particles based on~$S'$---corresponds in Euclidean field theory on a spin
manifold~$X$ to a section of a complex spin bundle over~$X$.  The details
depend on the dimension~$n$; see the text before~\thetag{2.5}.  For
simplicity we denote this bundle as $\Pi \SSC\otimes E\to X$, where $\SSC\to X$
is a complex spin bundle and $E\to X$ a complex vector bundle.\footnote{As
explained in~\S2 this is correct for example if $n\equiv3\pmod8$, but must be
modified for example if $n\equiv6\pmod8$.}  We can take $E$~to be trivial or
we can take it to be an arbitrary bundle with connection, in which case the
connection is a bosonic field in the theory.  The Euclidean
action\footnote{See~\cite{DF,\S7.4} for a derivation, though what is written
there should be modified in some dimensions, such as $n\equiv6\pmod8$.} is
the bilinear form~\thetag{2.6} based on the Riemannian Dirac operator on~$X$
coupled to~$E$.  Then the functional integral over the spinor field is
  $$ \pfaff \Dirac_E, \tag{A.25} $$
the pfaffian of the Dirac operator on spinors coupled to~$E$, as
in~\thetag{2.7}.  
 
For Rarita-Schwinger fields we proceed similarly using our first approach to
quantization in Minkowski spacetime.  Thus the Euclidean Rarita-Schwinger
field is a section of $\Pi \SSC\otimes E\otimes T_{\CC}X\to X$.  Since the
Euclidean version of the lagrangian~\thetag{A.10} is the action of a spinor
field coupled to~$E\otimes \TCX$, the partition function is
  $$ \pfaff \Dirac_{E\otimes \TCX}. \tag{A.26} $$
However, this corresponds to the physical theory with too many particles: the
desired spin~3/2 particles plus three extraneous collections of spin~1/2
particles; see~\thetag{A.12}.  Thus \thetag{A.26}~represents the product of
the desired partition function with the partition functions~\thetag{A.25} of
Euclidean spinor fields.  The partition function we seek, which corresponds
to the desired collection of spin~3/2 particles, is obtained by dividing out
the spinor partition functions.  Let $\tilde\Dirac_{\tilde E}$~denote the
Dirac operator on the Euclidean spinor bundle~$\tilde\SSC\otimes \tilde E$
which corresponds to the Lorentz spin representation~$S^*$; its relationship
to~$\Dirac_E$ depends on the dimension~$n$.  Then we define the
Rarita-Schwinger partition function to be
  $$ \frac{\pfaff \Dirac_{E\otimes \TCX}}{(\pfaff \Dirac_E)^2(\pfaff
     \tilde\Dirac_{\tilde E})}. \tag{A.27} $$
Finally, the Rarita-Schwinger anomaly is a tensor product of pfaffian line
bundles 
  $$ L_{RS} = (\Pfaff \Dirac_{E\otimes \TCX})\otimes (\Pfaff
     \Dirac_E)^{\otimes (-2)}\otimes (\Pfaff \tilde\Dirac_{\tilde
     E})^{\otimes (-1)}. \tag{A.28} $$

 \subhead \S{A.6}. Two Illustrative Examples
 \endsubhead

The minimal supergravity in $n=10$~dimensions occurs in the low energy
approximation to superstring theories with 16~supersymmetries, and it
contains a Rarita-Schwinger field.  The appropriate real spinor
representation~$S$ of the Lorentz spin group is the minimal one of
dimension~16; its dual~$S^*$ is an inequivalent representation.\footnote{In
dimensions $n\equiv 2,6\pmod8$ there are two inequivalent minimal real spinor
representations, whereas in other dimensions there is only one.}  In the
corresponding Euclidean version $\SSC\to X$ is a half-spinor bundle of the
spin 10-manifold~$X$ and $\tilde \SSC\to X$ is the opposite chirality
half-spinor bundle.  The vector bundle~$E$ is a trivial line bundle, so plays
no role.  The Dirac operator $\Dirac\:\Gamma (\SSC)\to \Gamma (\tilde\SSC)$ is
complex skew-adjoint relative to the duality pairing between~$\SSC$
and~$\tilde \SSC$.  The Dirac operator~$\tilde\Dirac$ in the denominator
of~\thetag{A.27} is the adjoint of~$\Dirac$ relative to the hermitian
structures on~$\Gamma (\SSC)$ and~$\Gamma (\tilde\SSC)$.  Its pfaffian line
bundle is the canonically the inverse\footnote{In the finite dimensional
model~\thetag{2.10} we are asserting that if $V\to T$ is a hermitian vector
bundle with connection, then $(\Det V^*)\otimes (\Det V)$ is canonically
isomorphic to the trivial line bundle with trivial metric and connection.} of
the pfaffian line bundle of~$\Dirac$, so that the anomaly~\thetag{A.28} in
this case simplifies to
  $$ (\Pfaff \Dirac_{\TCX})\otimes (\Pfaff \Dirac)^{\otimes
     (-1)}. \tag{A.29} $$
Formally, this is the pfaffian line bundle of Dirac coupled to~$\TCX-1$.
 
The second example is $n=11$~dimensional supergravity, which is the low
energy approximation to the (putative) M-theory.  Again $S$~is the minimal
real spinor representation of the Lorentz group, which in this case is unique
of dimension~32.  So~$ S^*\cong S$.  In the Euclidean version $\SSC\to X$ is
the rank~32 spinor bundle of the spin 11-manifold~$X$, and the Dirac
operators~$\Dirac $ and~$\tilde\Dirac $ agree.  The anomaly is then
  $$ (\Pfaff \Dirac_{\TCX})\otimes (\Pfaff \Dirac)^{\otimes
     (-3)}.  $$
Formally, this is the pfaffian line bundle of Dirac coupled to~$\TCX-3$.

\newpage
\widestnumber\key{SSSSSSSS}   % for widest bibliography name


\Refs\tenpoint

\ref
\key Ab      
\by L. Abrams 
\paper Two-dimensional topological quantum field theories and Frobenius
algebras 
\jour J. Knot Theory Ramifications 
\vol 5 
\yr 1996
\pages 569--587
\endref

\ref
\key AP      
\by  J. F. Adams,   S. B. Priddy
\paper Uniqueness of $BSO$
\jour Math. Proc. Cambridge Philos. Soc. 
\vol  80 
\yr  1976 
\pages 475--509
\endref

\ref 
\key AgW 
\by L. Alvarez-Gaum\'e, E. Witten \paper Gravitational anomalies \jour Nucl.
Phys. \vol B234 \yr 1983 \pages 269  
\endref

\ref
\key A1      
\by M. F. Atiyah
\paper Topological quantum field theory
\jour Publ. Math. Inst. Hautes Etudes Sci. (Paris)
\vol 68
\yr 1989
\pages 175--186                                      
\endref

\ref 
\key A2 
\by M. F. Atiyah 
\paper K-Theory Past and Present 
\finalinfo {\tt math.KT/001221}
\endref

\ref
\key AH      
\by M. F. Atiyah, F. Hirzebruch \paper Vector bundles and homogeneous spaces\jour Proc. Symp. Pure Math. \vol 3 \yr 1961 \pages 7--38 
\endref

\ref
\key AS1     
\by M. F. Atiyah, I. M. Singer \paper The index of elliptic operators. I \jour Ann. Math. \vol 87 \yr 1968 \pages 484--530
\endref

\ref
\key AS2     
\by M. F. Atiyah, I. M. Singer \paper The index of elliptic operators. IV \jour Ann. Math. \vol 93 \yr 1971 \pages 119--138
\endref

\ref
\key AS3     
\by M. F. Atiyah, I. M. Singer \paper The index of elliptic operators V\jour Ann. of Math. \vol 93 \yr 1971 \pages 139--149 
\endref

\ref
\key AS4     
\by M. F. Atiyah, I. M. Singer \paper Dirac operators coupled to vector potentials
\jour Proc. Nat. Acad. Sci. \vol 81 \yr 1984 \page 2597
\endref

\ref
\key ASe     
\by M. F. Atiyah, G. B.  Segal 
\paper  Exponential isomorphisms for $\lambda $-rings
\jour Quart. J. Math. Oxford Ser. 
\vol  22 
\yr  1971 
\pages 371--378
\endref

\ref
\key BW      
\by W. A. Bardeen, A. R. White 
\book Symposium on Anomalies, Geometry, Topology (March, 1985) 
\publ World Scientific 
\yr 1985 
\publaddr Singapore
\endref

\ref 
\key  BCMMS
\by P. Bouwknegt, A. L. Carey, V. Mathai, M. K. Murray,
D. Stevenson
\paper Twisted K-theory and K-theory of bundle gerbes 
\finalinfo hep-th/0106194
\endref

\ref
\key BF      
\by J. M. Bismut, D. S. Freed \paper The analysis of elliptic
families I: Metrics and connections on determinant bundles \jour Commun. Math.
Phys. \vol 106 \pages 159--176 \yr 1986
\moreref
\paper The analysis of elliptic families II:
Dirac operators, eta invariants, and the holonomy theorem of Witten \jour
Commun. Math. Phys. \vol 107 \yr 1986 \pages 103--163
\endref

\ref
\key BS      
\by A. Borel, J.-P. Serre \paper Le th\'eor\`eme de Riemann-Roch\jour Bull. Soc. Math. France \vol 86 \yr 1958 \pages 97--136 
\endref

\ref
\key B       
\by{Brylinski, Jean-Luc}
\book{Loop spaces, characteristic classes and geometric quantization}
\publ{Birkh\"auser Boston Inc.}
\publaddr{Boston, MA}
\endref

\ref 
\key CHS 
\by  C. G. Callan, Jr., J. A. Harvey, A. Strominger
\paper{Supersymmetric string solitons}
\inbook{String theory and quantum gravity (Trieste, 1991)}
\pages{208--244}
\publ{World Sci. Publishing}
\publaddr{River Edge, NJ}
\yr{1992} 
\finalinfo{\tt hep-th/9112030}
\endref

\ref
\key CY      
\by{Y.-K. E. Cheung, Z. Yin}
\paper{Anomalies, branes, and currents}
\jour{Nuclear Phys. B}
\vol{517}
\yr{1998}
\pages{69--91}
\finalinfo {\tt hep-th/9710206}
\endref


\ref
\key C       
\by J. Cheeger \paper $\eta$-invariants, the adiabatic  approximation, and conical singularities \jour Jour. Diff. Geo. \vol 26 \yr 1987 
\endref

\ref
\key De      
\by P. Deligne
\paper Notes on spinors
\pages 99--135
\eds{P. Deligne, P. Etingof, D. S. Freed, L. C. Jeffrey, D. Kazhdan,
J. W. Morgan, D. R. Morrison, E. Witten} 
 \yr{1999}
 \publaddr{Providence, RI}
 \bookinfo{2~volumes}
\inbook Quantum Fields and Strings:  A Course for Mathematicians
\publ American Mathematical Society 
\endref

\ref
\key DF      
\by P. Deligne, D. S. Freed 
\paper{Classical field theory}
\inbook{Quantum Fields and Strings: A Course for Mathematicians}
\eds{P. Deligne, P. Etingof, D. S. Freed, L. C. Jeffrey, D. Kazhdan,
J. W. Morgan, D. R. Morrison, E. Witten} 
 \publ{American Mathematical Society}
 \yr{1999}
 \publaddr{Providence, RI}
 \bookinfo{2~volumes}
\pages{137--225}
\endref


\ref 
\key DMW 
\by D.-E. Diaconescu, G. Moore, E. Witten
\paper E8 Gauge Theory, and a Derivation of K-Theory from M-Theory
\finalinfo {\tt hep-th/0005090}
\endref

\ref
\key D       
\by R. Dijkgraaf 
\paper A geometrical approach to two-dimensional conformal field theory
\paperinfo Ph.D. thesis 
\yr 1989
\endref

\ref
\key DK      
\by P. Donovan, M.  Karoubi 
\paper  Graded Brauer groups and $K$-theory with local coefficients
\jour Inst. Hautes \'Etudes Sci. Publ. Math. 
\vol 38 
\yr  1970 
\pages 5--25
\endref

\ref
\key Fyn
\by R. P. Feynman, R. B. Leighton, M. Sands
\book The Feynman Lectures on Physics
\bookinfo Volume~2
\publ Addison-Wesley
\publaddr Reading, Massachusetts
\yr 1964
\endref


\ref
\key F1      
\by D. S. Freed
\paper Higher algebraic structures and quantization
\jour Commun. Math. Phys.
\vol 159
\pages 343--398
\yr 1994
\finalinfo {\tt hep-th/9212115}
\endref

\ref
\key F2      
\by D. S. Freed
\paper Extended structures in topological quantum field theory
\inbook Quantum Topology
\eds L. H. Kauffman, R. A. Baadhio
\pages 162--173
\finalinfo {\tt hep-th/9306045}
\endref

\ref
\key F3      
\by D. S. Freed 
\paper The Verlinde algebra is twisted equivariant $K$-theory 
\jour Turkish J. Math. 
\vol 25 
\yr 2001 
\pages 159--167
\finalinfo {\tt math.RT/0101038}
\endref

\ref
\key F4      
\by D. S. Freed \paper On determinant line bundles \inbook Mathematical
Aspects of String Theory \bookinfo ed. S. T. Yau \publ World Scientific
Publishing \yr 1987 \pages 189--238
\endref

\ref
\key F5      
\by D. S. Freed
\paper Dirac charge quantization and generalized differential cohomology
\paperinfo{Papers Dedicated to Atiyah, Bott, Hirzebruch, Shapiro} 
\toappear 
\publ{International Press}
\finalinfo {\tt hep-th/0011220} 
\endref

\ref 
\key F6 
\by D. S. Freed 
\book{Five lectures on supersymmetry}
\publ{American Mathematical Society}
\publaddr{Providence, RI}
\yr{1999} 
\endref

\ref
\key FH      
\by D. S. Freed, M. J. Hopkins 
\paper On Ramond-Ramond fields and $K$-theory  
\jour J. High Energy Phys.  
\yr 2000 
\paperinfo Paper 44
\finalinfo {\tt hep-th/0002027}
\endref

\ref
\key FQ      
\by D. S. Freed, F. Quinn
\paper Chern-Simons theory with finite gauge group
\jour Commun. Math. Phys.
\yr 1993
\vol 156
\finalinfo{\tt hep-th/9111004}
\endref

\ref
\key FW      
\by D. S. Freed, E. Witten 
\paper Anomalies in String Theory with D-Branes 
\jour Asian J. Math 
\vol 3 
\yr 1999 
\pages 819--851
\finalinfo {\tt hep-th/9907189}
\endref

\ref  
\key GHM 
\by M. B. Green, J. A. Harvey, G. Moore 
\paper I-Brane Inflow and Anomalous Couplings on D-Branes
\jour Class. Quantum Grav. 
\vol 14 
\yr 1997 
\pages 47--52
\finalinfo{\tt hep-th/9605033}
\endref


\ref
\key GS      
\by M. B. Green, J. H. Schwarz 
\paper Anomaly cancellations in supersymmetric $D=10$ gauge theory and
superstring theory 
\jour Phys. Lett. B 
\yr 1984
\endref

\ref
\key H       
\by N. J. Hitchin 
\paper Lectures on special lagrangian submanifolds 
\inbook{Winter School on Mirror Symmetry, Vector Bundles and
              Lagrangian Submanifolds (Cambridge, MA, 1999)}
\pages{151--182}
\publ{Amer. Math. Soc.}
\publaddr{Providence, RI}
\yr{2001}
\finalinfo {\tt math.DG/9907034}
\endref

\ref 
\key d'H 
\by E. d'Hoker 
\paper String Theory 
\inbook{Quantum Fields and Strings: A Course for Mathematicians}
\eds{P. Deligne, P. Etingof, D. S. Freed, L. C. Jeffrey, D. Kazhdan,
J. W. Morgan, D. R. Morrison, E. Witten} 
 \publ{American Mathematical Society}
 \yr{1999}
 \publaddr{Providence, RI}
 \bookinfo{2~volumes}
\pages{807--1011}
\endref


\ref
\key HS      
\by M. J. Hopkins, I. M. Singer 
\paper Quadratic functions in geometry, topology, and M-theory 
\miscnote in preparation
\endref

\ref
\key L       
\paper Triangulations, categories and extended topological field theories
\by R. Lawrence         
\inbook Quantum Topology
\eds L. H. Kauffman, R. A. Baadhio
\pages 191--208
\publ World Scientific
\endref

\ref
\key Lo      
\by Lott, J.
\paper{$\RR/\ZZ$ index theory}
\jour{Comm. Anal. Geom.}
\vol{2}
\yr{1994}
\pages 279--311
\endref

\ref 
\key MM 
\by R. Minasian, G. Moore 
\paper $K$-theory and Ramond-Ramond charge
\jour J. High Energy Phys.
\yr 1998
\finalinfo no.~11, Paper 2, 7 pp, {\tt hep-th/9710230}
\endref


\ref
\key M       
\by G. Moore 
\paper Some comments on branes, $G$-flux, and $K$-theory. 
\jour Internat. J. Modern Phys. A 
\vol 16 
\yr 2001 
\pages 936--944 
\finalinfo {\tt hep-th/0012007}
\endref

\ref
\key MW      
\by G. Moore, E. Witten 
\paper Self-duality, Ramond-Ramond fields, and $K$-theory 
\jour J. High Energy Phys. 
\yr 2000 
\issue 5 
\paperinfo Paper 32
\finalinfo {\tt hep-th/9912279}
\endref

\ref 
\key P1 
\by{J. Polchinski}
\book{String theory. {V}ol. {I}{I}}
\publ{Cambridge University Press}
\publaddr{Cambridge}
\yr{1998} 
\endref 
 
\ref 
\key P2 
\by{J. Polchinski}
\paper{Lectures on {D}-branes}
\inbook{Fields, strings and duality (Boulder, CO, 1996)}
\pages{293--356}
\publ{World Sci. Publishing}
\publaddr{River Edge, NJ}
\yr{1997} 
\finalinfo{\tt hep-th/9611050}
\endref


\ref
\key Qn      
\by D. Quillen \paper Determinants of Cauchy-Riemann operators over a Riemann surface \jour Funk. Anal. iprilozen \vol 19 \yr 1985 \pages 37
\endref

\ref
\key Q       
\by F. Quinn
\paper Lectures on axiomatic topological quantum field theory 
\inbook Geometry and quantum field theory (Park City, UT, 1991) 
\pages 323--453
\bookinfo IAS/Park City Math. Ser., 1 
\publ Amer. Math. Soc. 
\publaddr Providence, RI 
\yr 1995
\endref

\ref
\key RS      
\by D. B. Ray, I. M. Singer \paper $R$-torsion and the laplacian on Riemannian
manifolds \jour Adv. Math. \vol 7 \yr 1971 \pages 145--210 
\endref

\ref
\key R       
\by J. Rosenberg 
\paper Continuous trace algebras from the bundle theoretic point of view 
\jour Jour. Austr. Math. Soc. 
\vol 47 
\yr 1989 
\pages 368--381
\endref

\ref
\key Sa      
\by S. Sawin 
\jour J. Math. Phys. 
\vol 36 
\yr 1995 
\pages 6673--6680
\paper Direct sum decompositions and indecomposable TQFTs
\finalinfo{\tt q-alg/9505026}
\endref

\ref
\key S1      
\by G.  Segal 
\paper Categories and cohomology theories
\jour Topology 
\vol  13 
\yr  1974 
\pages 293--312
\endref

\ref
\key S2      
\by G. Segal
\paper The definition of conformal field theory
\miscnote preprint
\endref

\ref 
\key Se 
\by A. Sen 
\paper Tachyon Condensation on the Brane Antibrane System 
\jour JHEP 9808 (1998) 012 
\finalinfo{\tt hep-th/9805170}
\endref


\ref
\key Si1     
\by I. M. Singer 
\paper Families of Dirac operators with applications to physics.  
\paperinfo The mathematical heritage of \'Elie Cartan (Lyon, 1984) 
\jour Astérisque
\yr 1985 
\pages 323--340
\endref

\ref
\key Si2     
\by I. M. Singer \paper The $\eta $-invariant and the index 
\inbook Mathematical
Aspects of String Theory \bookinfo ed. S. T. Yau \publ World Scientific
Publishing \yr 1987 \pages
\endref

\ref
\key T1      
\by V. G. Turaev
\book Quantum Invariants of Knots and 3-Manifolds
\publ Walter de Gruyter
\publaddr Berlin
\yr 1994
\endref

\ref
\key T2      
\by V. Turaev
\paper Homotopy field theory in dimension 3 and crossed group-categories
\finalinfo {\tt math.GT/0005291}
\endref

\ref
\key Wa      
\by K. Walker
\paper On Witten's 3-manifold invariants
\miscnote preprint, 1991
\endref

\ref
\key W1      
\by E. Witten
\paper Quantum field theory and the Jones polynomial
\jour Commun. Math. Phys.
\vol 121
\yr 1989
\endref

\ref
\key W2      
\by E. Witten
\paper World-sheet corrections via D-instantons 
\jour J. High Energy Phys.  
\vol 2000 
\issue 2
\paperinfo Paper 30
\finalinfo {\tt hep-th/9907041}
\endref

\ref
\key W3      
\by E. Witten \paper Global gravitational anomalies \jour Commun. Math. Phys.
\vol 100 \yr 1985 \pages 197--229 
\endref

\ref
\key W4      
\finalinfo {\tt hep-th/9810188}
\by E. Witten
\paper D-branes and $K$-theory
\jour J. High Energy Phys. 
\vol 1998 
\issue 12 
\paperinfo Paper 19
\endref

\ref 
\key W5 
\by E. Witten
\paper Five-brane effective action in $M$-theory
\jour J. Geom. Phys. 
\vol 22  
\yr 1997 
\pages 103--133
\finalinfo {\tt hep-th/9610234}
\endref



\endRefs
\enddocument